\documentclass[11pt,american]{article}
\usepackage{amsmath,amsfonts}
\usepackage{amsthm}
\usepackage{times}
\usepackage{libertineRoman}
\usepackage{biolinum}

\usepackage[libertine]{newtxmath}
\usepackage[T1]{fontenc}
\usepackage[latin9]{inputenc}
\usepackage{geometry}
\usepackage{color}
\usepackage{refstyle}
\usepackage{textcomp}
\usepackage{cancel}
\usepackage{graphicx}
\PassOptionsToPackage{normalem}{ulem}
\usepackage{ulem}
\usepackage{authblk}

\makeatletter

\AtBeginDocument{\providecommand\Figref[1]{\ref{Fig:#1}}}
\AtBeginDocument{\providecommand\Eqref[1]{\ref{Eq:#1}}}
\AtBeginDocument{\providecommand\Thmref[1]{\ref{Thm:#1}}}
\AtBeginDocument{\providecommand\Secref[1]{\ref{Sec:#1}}}
\AtBeginDocument{\providecommand\Defref[1]{\ref{Def:#1}}}
\AtBeginDocument{}
\AtBeginDocument{\providecommand\Propref[1]{\ref{Prop:#1}}}
\AtBeginDocument{}
\AtBeginDocument{}
\AtBeginDocument{\providecommand\Lemref[1]{\ref{Lem:#1}}}
\RS@ifundefined{subsecref}
  {\newref{subsec}{name = \RSsectxt}}
  {}
\RS@ifundefined{thmref}
  {\def\RSthmtxt{theorem~}\newref{thm}{name = \RSthmtxt}}
  {}
\RS@ifundefined{lemref}
  {\def\RSlemtxt{lemma~}\newref{lem}{name = \RSlemtxt}}
  {}

\theoremstyle{plain}
\newtheorem{thm}{\protect\theoremname}
\theoremstyle{definition}
\newtheorem{defn}[thm]{\protect\definitionname}
\theoremstyle{plain}
\newtheorem{lem}[thm]{\protect\lemmaname}
\theoremstyle{definition}

\theoremstyle{remark}

\theoremstyle{plain}

\theoremstyle{plain}
\newtheorem{prop}[thm]{\protect\propositionname}
\theoremstyle{plain}
\newtheorem*{thm*}{\protect\theoremname}

\usepackage{tikz}
\usetikzlibrary{calc}

\usepackage{enumitem}
\setlistdepth{20}
\renewlist{itemize}{itemize}{20}
\setlist[itemize]{label=$\cdot$}

\setlist[itemize,1]{label=\textbullet}
\setlist[itemize,2]{label=--}
\setlist[itemize,3]{label=*}
\setlist[itemize,4]{label=$\circ$}
\setlist[itemize,5]{label=$\square$}

\usepackage{hyperref}

\hypersetup{colorlinks=true,urlcolor=blue}

\usepackage[super]{nth}

\usepackage{braket}

\usepackage{flafter}

\usepackage{psvectorian}

\newref{thm}{name=theorem~,Name=Theorem~,names=theorems~,Names=Theorems~}
\newref{def}{name=definition~,Name=Definition~,names=definitions~,Names=Definitions~}
\newref{cor}{name=corollary~,Name=Corollary~,names=corollaries~,Names=Corollaries~}
\newref{lem}{name=lemma~,Name=Lemma~,names=lemmas~,Names=Lemmas~}
\newref{rem}{name=remark~,Name=Remark~,names=remarks~,Names=Remarks~}
\newref{exa}{name=example~,Name=Example~,names=examples~,Names=Examples~}
\newref{sec}{name=section~,Name=Section~,names=sections~,Names=Sections~}
\newref{subsec}{name=subsection~,Name=Subsection~,names=subsections~,Names=Subsections~}
\newref{prop}{name=proposition~,Name=Proposition~,names=propositions~,Names=Propositions~}

\usepackage{color}
\definecolor{purple}{RGB}{120,20,120}

\newcommand{\overbar}[1]{\mkern 1.75mu\overline{\mkern-1.75mu#1\mkern-1.75mu}\mkern 1.75mu}

\@ifundefined{showcaptionsetup}{}{%
 \PassOptionsToPackage{caption=false}{subfig}}
\usepackage{subfig}
\makeatother

\usepackage{babel}
\usepackage[style=numeric,maxbibnames=8,maxcitenames=8,maxnames=8]{biblatex}

\providecommand{\corollaryname}{Corollary}
\providecommand{\definitionname}{Definition}
\providecommand{\examplename}{Example}
\providecommand{\lemmaname}{Lemma}
\providecommand{\propositionname}{Proposition}
\providecommand{\remarkname}{Remark}
\providecommand{\theoremname}{Theorem}
\begin{document}
\title{Analytic quantum weak coin flipping protocols with arbitrarily
small bias}
\author{Atul Singh Arora\thanks{aarora@ulb.ac.be}}
\author{J\'er\'emie Roland\thanks{jroland@ulb.ac.be}}
\author{Chrysoula Vlachou\thanks{cvlachou@ulb.ac.be}}

\affil{\small{}Universit\'e libre de Bruxelles, Belgium}
\date{13 July 2020}
\maketitle

\begin{abstract}
 Weak coin flipping (WCF) is a fundamental cryptographic primitive for two-party secure computation, where two distrustful parties need to remotely establish a shared random bit whilst having opposite preferred outcomes. It is the strongest known primitive with arbitrarily close to perfect security quantumly while classically, its security is completely compromised (unless one makes further assumptions, such as computational hardness).  A WCF protocol is said to have bias $\epsilon$ if neither party can force their preferred outcome with probability greater than $1/2+\epsilon$. Classical WCF protocols are shown to have bias $1/2$, i.e., a cheating party can always force their preferred outcome. On the other hand, there exist quantum WCF protocols with arbitrarily small bias, as Mochon showed in his seminal work in 2007 [arXiv:0711.4114]. In particular, he proved the existence of a family of WCF protocols approaching bias $\epsilon (k)=1/(4k+2)$ for arbitrarily large $k$ and proposed a protocol with bias $1/6$. Last year, Arora, Roland and Weis presented a protocol with bias $1/10$ and to go below this bias, they designed an algorithm that \emph{numerically} constructs unitary matrices corresponding to WCF protocols with arbitrarily small bias [STOC'19, p.205-216]. In this work, we present new techniques which yield a fully analytical construction of WCF protocols with bias arbitrarily close to zero, thus achieving a solution that has been missing for more than a decade. Furthermore, our new techniques lead to a simplified proof of existence of WCF protocols by circumventing the non-constructive part of Mochon's proof. As an example, we illustrate the construction of a WCF protocol with bias $1/14$.

\end{abstract}

\vfill

\pagebreak{}

\global\long\def\diag{\text{diag}}%

\global\long\def\tr{\text{tr}}%

\global\long\def\hc{\text{h.c.}}%

\global\long\def\prob{\text{Prob}}%

\global\long\def\Prob{\text{Prob}}%

\global\long\def\ol#1{\overbar{#1}}%

\global\long\def\ob#1{\overbar{#1}}%

\global\long\def\ket#1{\left|#1\right\rangle }%

\global\long\def\bra#1{\left\langle #1\right|}%

\global\long\def\minmax#1#2{\left(#1,#2\right)}%

\section{Introduction}

Coin flipping (CF), introduced by Blum \cite{Blum:1983:CFT:1008908.1008911}, is an important
cryptographic primitive which permits two distrustful parties to remotely generate an unbiased random bit in spite of
the fact that one of them might be dishonest and try to force a
specific outcome. Like bit commitment (BC)  and oblivious transfer (OT), it is a basic primitive for secure 2-party computation, a special case of secure multi-party computation, where the parties need to jointly compute a function on their inputs while keeping these inputs private. In the classical scenario, these primitives are shown to be computationally secure, and without extra assumptions (e.g. computational hardness) a dishonest party can always cheat perfectly \cite{Cleve86}.
Moving to the quantum scenario, BC and OT protocols have a non-zero lower bound on their bias \cite{CK11,Chailloux2013}; achieving perfect security is not possible, but still they perform better than their classical counterparts without computational hardness assumptions. 
The two distinct variants of CF, namely strong CF (SCF)
and weak CF (WCF), behave differently in the quantum scenario. In SCF
the desired outcome of each party is not known a priori, i.e., none
of the parties know beforehand whether the other prefers outcome $0$
or $1$. Just like for quantum BC and OT, there is a lower bound on the bias of SCF protocols \cite{LO1998177,Kitaev03}. The best known explicit quantum SCF protocols had bias $\frac{1}{4}$ \cite{Ambainis04b,NS03,KerenidisNayak04}. 
For a quantum WCF protocol though, where the preferred outcome of
each party is known, the situation is different. In his seminal work,
Mochon \cite{Mochon07} proved the existence of a family of WCF protocols
achieving arbitrarily close to zero bias. This established WCF to be the strongest known  secure 2-party computation primitive which has arbitrarily close to perfect security in the quantum setting while being completely insecure classically (without making further assumptions). Moreover, Kerenidis and Chailloux showed that perfect WCF can be used as a block box to obtain the optimal protocols for quantum SCF and BC \cite{CK09,CK11}, i.e. the protocols with the lowest possible bias $\frac{1}{\sqrt{2}} - \frac{1}{2}$, therefore Mochon's result is highly relevant for the whole area of quantum secure 2-party (and multi-party) computation.
However, his proof was not constructive and the proposal of an explicit protocol with almost zero bias was left as an open problem, while only an explicit protocol with bias $\frac{1}{6}$ was presented. In fact, first, a WCF protocol with bias $\frac{1}{\sqrt{2}} - \frac{1}{2}$ was reported \cite{Spekkens2002}, which incidentally matched the \emph{lower bound} for the bias of \emph{SCF} protocols, undermining even the existence of better WCF protocols and the distinction between them. Later, Mochon's lengthy and highly technical proof was verified and simplified \cite{ACG+14}, but still a protocol with bias below $\frac{1}{6}$ was missing. Last year Arora, Roland
and Weis proposed an explicit protocol with bias $\frac{1}{10}$, and designed an algorithm that can \textit{numerically} construct unitary matrices corresponding to protocols with arbitrarily small bias \cite{Arora2019}.
In the present work, we report the \textit{analytical} solution
to the WCF problem, by determining the unitary matrices that constitute WCF protocols with arbitrarily small bias.

\section{Background and overview of the result\label{sec:Preliminaries}}

A quantum WCF protocol can be described
as follows: the two parties, say A and B, are located in different
places and, besides their local register, they also have a register that they can exchange, called the message register. At
each round, the party that holds the message register
can apply a local unitary on it and on their local register. After
a number of rounds, the parties perform a final measurement on their local registers,
and the outcome determines the winner: A wins on outcome $0$, while B wins on outcome $1$. If both parties are honest and  follow the protocol, they have equal probabilities of
winning $P_{A}=P_{B}=1/2$. If one of the parties
is cheating and tries
to force the other player to output their desired outcome, then
their probability of winning is, in general, greater.
We denote this probability by $P_{A}^{*}$ for  A being dishonest
and $P_{B}^{*}$ for dishonest B. Let $\epsilon\ge0$ be the smallest
number such that both $P_{A}^{*}$ and $P_B^*$ are upper bounded by $\frac{1}{2}+\epsilon$. Then we say that the protocol has \emph{bias} $\epsilon$.\footnote{The case where both A and B are dishonest does not depend on the description of the protocol since neither is following it.} To calculate
$P_{A/B}^{*}$ one can write a semi-definite program (SDP) that maximizes
this cheating probability, given that the honest party
follows the protocol. Using the SDP duality, this maximization
problem can be written as a minimization problem over the respective
dual variables $Z_{A/B}$. However, the above holds given that we already have a protocol. Therefore, a new framework
is needed, permitting us to find both the protocol and its bias.

A ground-breaking idea was provided by Kitaev (as Mochon describes in \cite{Mochon07}), who transformed these
SDPs into the so-called \textit{time-dependent point
games} (TDPG). A TDPG is a sequence of frames that include
points on the positive quadrant of the $x-y$ plane with a probability weight assigned
to each point. The TDPGs that
we consider are determined by specific initial and
final configurations and there are rules on how to move from one frame to the next. The initial frame has
two points with coordinates $\llbracket 0,1 \rrbracket $ and $\llbracket 1,0 \rrbracket $ and probability
weight $1/2$ each, while the final frame we want to obtain has only one
point at $\llbracket \beta,\alpha \rrbracket $ with probability weight $1$.
Consider one
frame, and restrict to the set of points along a horizontal line, i.e.
points with the same $y$ coordinate. We denote the $x-$coordinates of the $i$th
such point by $x_{g_{i}}$ and the respective probability weight
by $p_{g_{i}}$, with $i\in\{1,2\dots n_{g}\}$. In the
subsequent frame, restrict again to a set of points with the same
$y$ coordinate as before. Let the $x-$coordinates of the $i$th such point
be $x_{h_{i}}$ and the respective probability weight be $p_{h_{i}}$,
with $i\in\{1,2\dots n_{h}\}$. The
rules for transitioning between subsequent frames can be written as follows:

\begin{equation}
\sum_{i=1}^{n_{g}}p_{g_{i}}  =\sum_{i=1}^{n_{h}}p_{h_{i}}\ \ \ \ \ \ \text{and}\ \ \ \ \ 
\sum_{i=1}^{n_{g}}\frac{\lambda x_{g_{i}}}{\lambda+x_{g_{i}}}p_{g_{i}} \le\sum_{i=1}^{n_{h}}\frac{\lambda x_{h_{i}}}{\lambda+x_{h_{i}}}p_{h_{i}}, \ \ \ \ \forall\lambda>0.\label{eq:const}
\end{equation}
Analogous rules exist for moving points along  vertical lines.
Some examples of such permitted moves are the
\textit{raises}, where we move a point along a horizontal or vertical line by increasing its coordinate, the \textit{splits}, where we split a point into several others, and the \textit{merges}, where we merge several points into a single point.

It was shown that for any TDPG with transitions
respecting \Eqref{const}, there exists a WCF protocol
with cheating probabilities $P_{A}^{*}=\alpha+\delta$ and $P_{B}^{*}=\beta+\delta$, where $\delta$ can be made arbitrarily small. The converse also holds.
Thus, the initial task of finding a protocol and solving the associated
SDPs minimising $P_{A/B}^{*}$, is reduced to finding a TDPG such that the point $\llbracket \beta,\alpha \rrbracket $
of the final frame is as close to $\llbracket \frac{1}{2},\frac{1}{2} \rrbracket $ as
possible, corresponding to the zero-bias case. 
These TDPGs are called \emph{expressible
by matrices (EBM)} point games, and they are defined below.
 
\begin{defn}
Let $Z\geq0$ be a Hermitian matrix\footnote{This matrix inequality denotes that $Z$ is a positive semi-definite matrix.} and $\Pi^{[z]}$ be the projector on the eigenspace
of the eigenvalue $z$ of $Z$. Let $\ket{\psi}$
be a vector (not necessarily normalised), and define the finitely supported
function $\text{Prob}[Z,\ket{\psi}]:\left[0,\infty\right)\rightarrow\left[0,\infty\right)$
as 
\[
\text{Prob}[Z,\ket{\psi}](z)=\begin{cases}
\langle\psi|\Pi^{[z]}|\psi\rangle & \text{if }z\in\text{spectrum}(Z)\\
0 & \text{otherwise}.
\end{cases}
\]
Let $g,h:\left[0,\infty\right)\rightarrow[0,\infty)$ be two finitely supported functions. The line transition $g\rightarrow h$ is called \emph{EBM} if there exist two matrices $0\leq G\leq H$ and
a vector $\ket{\psi}$, such that 
$g=\text{Prob}[G,\ket{\psi}]\ \text{and}\ h=\text{Prob}[H,\ket{\psi}].$
\label{def:EBM_function}
\end{defn}

\begin{defn}
Let $g,h:[0,\infty)\times[0,\infty)\rightarrow[0,\infty)$
be two finitely supported functions. The transition $g\rightarrow h$ is called an
\begin{itemize}
\item \emph{EBM horizontal transition} if for all $y\in\left[0,\infty\right),g(\cdot,y)\rightarrow h(\cdot,y)$
is an EBM line transition, and 
\item \emph{EBM vertical transition} if for all $x\in\left[0,\infty\right),g(x,\cdot)\rightarrow h(x,\cdot)$
is an EBM line transition. 
\end{itemize}
\end{defn}

\begin{defn}
An \emph{EBM point game} is a sequence of finitely supported functions\footnote{As explained further in \Secref{Mochon's-Assignments}, $\llbracket a,b \rrbracket (x,y) := \delta_{a,x} \delta_{b,y}$ where $\delta_{r,s}$ is the Kronecker Delta.}
$\{g_{0},g_{1},\ldots,g_{n}\}$, such that
\begin{itemize}
\item $g_{0}=\frac{1}{2}\llbracket 0,1 \rrbracket+\frac{1}{2}\llbracket 1,0 \rrbracket$ and $g_{n}=1\llbracket \beta,\alpha \rrbracket$ for some $\alpha,\beta\in [0,1]$, 
\item for all even (odd) $i$ the transition $g_{i}\rightarrow g_{i+1}$ is an
EBM vertical (horizontal) transition. 

\end{itemize}
\end{defn}

In order to verify that a transition is EBM one has to check conditions
involving matrices, thus the problem remains hard and yet another reduction is needed. For an EBM transition
$g\rightarrow h$, one can consider the corresponding finitely supported \emph{EBM function} to be $h-g$. The set of EBM
functions is shown to be the same (up to the closures) as the set of the so-called
\textit{valid} functions. We omit both the definition
of a valid function and the proof that the two sets are
same, as they have been presented in previous works \cite{Mochon07,ACG+14}.
We only highlight that checking if a transition is
EBM is equivalent to verifying the validity of a suitably constructed
function which is an easier task.

Mochon, following the above reductions, proved
the existence of a WCF protocol with arbitrarily small bias,
by proposing a suitable family of point games with valid transitions \cite{Mochon07}. This family is parametrised by an arbitrary integer
$k\geq1$ that specifies the bias $\epsilon=\frac{1}{4k+2}$. More precisely, $2k$ is the number of points involved in the main move of the point game. He constructed a protocol with bias $\frac{1}{6}$, but he left as an open problem the construction of a protocol with almost zero bias. This problem has remained open since then, as
translating the point game into a sequence of unitaries
determining the protocol is, indeed, not easy. A step forward was
recently taken in \cite{Arora2019}, where a framework, TDPG-to-Explicit-protocol Framework (TEF) was introduced, which
allows the conversion of TDPGs into WCF protocols, granted that
unitaries associated with the valid functions used in the games can
be found. More precisely, if a unitary matrix $O$ acting on $\text{span}\{\ket{g_1}, \ket{g_2},\ldots,\ket{h_1}, \ket{h_2},\ldots\}$, and satisfying the constraints
\begin{equation}
O\ket{v}=\ket{w}\ \ \ \text{and}\ \ \ \sum_{i=1}^{n_h} x_{h_i}\ket{h_i}\bra{h_i}-\sum_{i=1}^{n_g} x_{g_i}E_h O\ket{g_i}\bra{g_i}O^{\dagger}E_h\geq 0,
\label{eq:TEF_const}
\end{equation}
can be found for every transition of a TDPG, then an explicit WCF protocol
with the corresponding bias can be obtained using the TEF. The vectors $\left\{\{\ket{g_i}\}_{i=1}^{n_g}, \{\ket{h_i}_{i=1}^{n_h}\}\right\}$ are orthonormal and $E_h$ is a projection on $\text{span}\{\ket{h_i}\}$.
Furthermore, $x_{g_i} \text{ and } x_{h_i}$ are the coordinates of the points of the initial and final frame, respectively, of the line transitions, and $p_{g_i} \text{ and } p_{h_i}$ their corresponding probability weights (see also \Eqref{const}). Note that there exist $n_g$ and $n_h$ points in the initial and final frame, respectively. Finally, $\ket{v}:=\sum_i \sqrt{p_{g_i}}\ket{g_i}/\sqrt{\sum_i p_{g_i}}$ and $\ket{w}:=\sum_i \sqrt{p_{h_i}}\ket{h_i}/\sqrt{\sum_i p_{h_i}}$. In fact the set of transitions which satisfy \Eqref{TEF_const} is the same (up to the closures) as the set of valid/EBM transitions (see Appendix \ref{app:sum-of-valid-functions}). Using a perturbative method in conjunction with the TEF, the authors in \cite{Arora2019} \emph{analytically} constructed a protocol with bias $\frac{1}{10}$, and to go below this bias they used tools from geometry, and designed the so-called
\emph{elliptic monotone align} algorithm, that \emph{numerically}
finds the matrices determining a protocol with arbitrarily small bias.

In the present work, we \emph{analytically} construct explicit WCF protocols with arbitrarily small bias, and to this end, we consider the class of valid functions that Mochon uses in his family
of point games approaching bias $\frac{1}{4k+2}$ for arbitrary integers $k\geq 1$. We refer to these valid functions as \emph{$f$-assignments}, and when
$f$ is a monomial, we call them \emph{monomial assignments}. We chose the term assignment to reflect the fact that these functions are assigning the appropriate probability weights to the points of the TDPGs. If we are able to construct unitaries satisfying \Eqref{TEF_const} with respect to the $f-$assignments of Mochon's TDPGs with bias $\epsilon\rightarrow 0\ \ (\text{i.e. for }k\rightarrow \infty)$, we have effectively solved our problem, since the aforementioned TEF enables the conversion of TDPGs to WCF protocols. We start by noticing that an even weaker condition is sufficient: suppose that a valid/EBM function can be written as a sum of valid/EBM functions; to obtain the protocol, it suffices to find unitaries corresponding to each valid function that appears in this sum (see Appendix \ref{app:sum-of-valid-functions}). We then solve the monomial assignments, i.e. we give formulae for the unitaries
corresponding to monomial assignments, and show that they indeed satisfy \Eqref{TEF_const}, obtaining, thus, an effective solution to the $f$-assignment, as summarised in our main result, \Thmref{Main}. 
Our approach, in addition to yielding  analytic WCF protocols with vanishing bias, has a feature that we would like to emphasize here. The reduction of the problem from EBM to valid functions is pivotal in the construction of Mochon's point game \cite{Mochon07}. However, we can bypass this reduction
and directly construct a WCF protocol
once the matrices $O$, corresponding to the (effective) solutions to the transitions of the point game, which satisfy \Eqref{TEF_const} are known. By means of the TEF we can prove that this protocol has the same bias as the point game. Therefore, our approach is simpler than the previous ones, as it avoids the aforementioned---quite technical---reduction.
Finally, in \cite{Arora2018,Arora2019} it was shown that functions expressible by \emph{real} matrices (EBRM) are sufficient for obtaining the solution,\footnote{This permitted the use of a geometric approach to achieve the numerical solution.} therefore from now on we restrict to orthogonal matrices.

\section{$f-$assignments and their properties\label{sec:Mochon's-Assignments}}

We write finitely supported functions
$t$ in two ways: (1) as $t=\sum_{i=1}^{n}p_{i}\llbracket x_{i}\rrbracket,$
where $|p_{i}|>0$ for all $i\in\{1,2\dots n\}$, and
$x_{i}\neq x_{j}$ for $i\neq j$, and (2) as $t=\sum_{i=1}^{n_{h}}p_{h_{i}}\llbracket x_{h_{i}}\rrbracket -\sum_{i=1}^{n_{g}}p_{g_{i}}\llbracket x_{g_{i}}\rrbracket $,
where $p_{h_{i}}, p_{g_{i}}>0$ and  $x_{h_{i}}, x_{g_{i}}$ are all distinct. By $\llbracket x_i \rrbracket$ we represent a point with coordinate $x_i$. More concretely, we have $\llbracket a \rrbracket (x)=\delta_{a,x}$, where $\delta_{a,x}$ is the Kronecker delta.
\begin{defn}[$f$-assignments]
 Given a set of real coordinates $0\le x_{1}<x_{2}\dots<x_{n}$ and a
polynomial of degree at most $n-2$ satisfying $f(-\lambda)\ge0$
for all $\lambda\ge0$, an \emph{$f$-assignment} is given by
the function
\[
t=\sum_{i=1}^{n}\underbrace{\frac{-f(x_{i})}{\prod_{j\neq i}(x_{j}-x_{i})}}_{=:p_{i}}\llbracket x_{i}\rrbracket =h-g,
\]
where $h$ contains the positive
part of $t$ and $g$ the negative part (without any common support),
viz. $h=\sum_{i:p_{i}>0}p_{i}\llbracket x_{i}\rrbracket $
and $g=\sum_{i:p_{i}<0}\left(-p_{i}\right)\llbracket x_{i}\rrbracket $.
\label{def:f_assignment-f_0_assignment-balanced-m_kmonomial-aligned}
\begin{itemize}

\item We say an assignment is \emph{balanced} if the number of points with
negative weights, $p_{i}<0$, equals the number of points with positive
weights, $p_{i}>0$. We say an assignment is \emph{unbalanced} if
it is not balanced.
\item When $f$ is a monomial, viz. has the form $f(x)=cx^{q}$, where $c>0$ and $q\geq0$, we call
the assignment a \emph{monomial assignment}. For $q=0$, we call the assignment the \emph{zeroth assignment}.
\item We say that a monomial assignment is \emph{aligned} if the degree of the monomial is an even number ($q=2(b-1),\ b\in\mathbb{N}$). We say that a monomial assignment is \emph{misaligned} if it is not aligned.
\end{itemize}
\end{defn}
In the definition above the coordinates are real non-negative numbers, but in the next sections where we present the solutions, we consider the coordinates to be strictly
positive. However, this is not really a restriction, because any
$f$-assignment with a zero coordinate can be expressed as an $f$-assignment
with strictly positive coordinates, in such a way that both have the
same solution (see \Lemref{OriginIssueHandled} in Appendix \ref{app:lemmas}).
\begin{defn}[ (Effectively) Solving an assignment]
Given a finitely supported function $t=\sum_{i=1}^{n_{h}}p_{h_{i}}\llbracket x_{h_{i}}\rrbracket -\sum_{i=1}^{n_{g}}p_{g_{i}}\llbracket x_{g_{i}}\rrbracket $
and an orthonormal basis $\left\{ \left|g_{1}\right\rangle ,\left|g_{2}\right\rangle \dots\left|g_{n_{g}}\right\rangle ,\left|h_{1}\right\rangle ,\left|h_{2}\right\rangle \dots\left|h_{n_{h}}\right\rangle \right\}, $
we say that an orthogonal matrix $O$ \emph{solves} $t$ if $O$ satisfies
the following: $O\left|v\right\rangle =\left|w\right\rangle $ and
$X_{h}\ge E_h OX_{g}O^{T}E_h$, where $\left|v\right\rangle =\sum_{i=1}^{n_{g}}\sqrt{p_{g_{i}}}\left|g_{i}\right\rangle $,
$\left|w\right\rangle =\sum_{i=1}^{n_{h}}\sqrt{p_{h_{i}}}\left|h_{i}\right\rangle $,
$X_{h}=\sum_{i=1}^{n_{h}}x_{h_{i}}\left|h_{i}\right\rangle \left\langle h_{i}\right|$,
$X_{g}=\sum_{i=1}^{n_{g}}x_{g_{i}}\left|g_{i}\right\rangle \left\langle g_{i}\right|$
and $E_{h}=\sum_{i=1}^{n_{h}}\left|h_{i}\right\rangle \left\langle h_{i}\right|$. Moreover, we say that $t $ has an \emph{effective solution} if $t=\sum_{i\in I}t'_i$ and $t'_i$ has a solution for all $i\in I$, where $I$ is a finite set.\label{def:solvinassign}
\end{defn}

    In Section \ref{sec:Preliminaries}, we claimed that in order to construct a WCF protocol with vanishing bias it suffices to obtain effective solutions to $f-$assignments. In particular, it suffices to express each $f-$assignment as a sum of monomial assignments and find the orthogonal matrices solving each monomial assignment appearing in the sum. In Appendix \ref{app:sum-of-valid-functions} we explain why this claim holds, and in \Lemref{generalMonomialDecomposition} below we show how an $f$-assignment\footnote{with real and non-negative roots,} can be trivially expressed as a sum of monomial assignments.

\begin{lem}[$f$-assignment as a sum of monomials]
  Consider a set of real coordinates\footnote{The restriction on the number of roots is justified by the forthcoming use of the $f-$assignment.} satisfying $0\le x_{1}<x_{2}\dots<x_{n}$
 and let $f(x)=(r_{1}-x)(r_{2}-x)\dots(r_{k}-x)$ where $k\le n-2$.
 Let $t=\sum_{i=1}^{n}p_{i}\left\llbracket x_{i}\right\rrbracket $
 be the corresponding $f$-assignment. Then 
 \[
 t=\sum_{l=0}^{k}\alpha_{l}\left(\sum_{i=1}^{n}\frac{-(-x_{i})^{l}}{\prod_{j\neq i}(x_{j}-x_{i})}\left\llbracket x_{i}\right\rrbracket \right),
 \]
 where $\alpha_{l}\ge0$. More precisely, $\alpha_{l}$ is the coefficient of
 $(-x)^{l}$ in $f(x)$.\label{lem:generalMonomialDecomposition}
 \end{lem}

In the following sections we present the orthogonal matrices solving the four possible types of monomial assignments, namely balanced/unbalanced and aligned/misaligned (see \Defref{f_assignment-f_0_assignment-balanced-m_kmonomial-aligned}).

\section{Solution to the  zeroth assignment}

In this section we present the solution for the simplest monomial assignment, which we call the \emph{zeroth} assignment, since $f(x)=(-x)^{0}$. We start with the orthogonal matrices solving the balanced case, and prove their correctness. Henceforth, we use \emph{h.c.} to denote the Hermitian conjugate.

\begin{prop}[Solution to balanced zeroth assignments]
 Let $t=\sum_{i=1}^{n}p_{h_{i}}\llbracket x_{h_{i}}\rrbracket -\sum_{i=1}^{n}p_{g_{i}}\llbracket x_{g_{i}}\rrbracket$ be a zeroth assignment over $0<x_{1}<x_{2}\dots<x_{2n}$, $\left\{ \left|h_{1}\right\rangle ,\left|h_{2}\right\rangle \dots\left|h_{n}\right\rangle ,\left|g_{1}\right\rangle ,\left|g_{2}\right\rangle \dots\left|g_{n}\right\rangle \right\} $
be an orthonormal basis, $E_{h}:=\sum_{i=1}^{n}\left|h_{i}\right\rangle \left\langle h_{i}\right|$ be a subspace projector,
and finally let

\[
X_{h}:=\sum_{i=1}^{n}x_{h_{i}}\left|h_{i}\right\rangle \left\langle h_{i}\right|\doteq\diag(x_{h_{1}},x_{h_{2}}\dots x_{h_{n}},\underbrace{0,0\dots0}_{n\text{-zeros}}),
\]
\[
X_{g}:=\sum_{i=1}^{n}x_{g_{i}}\left|g_{i}\right\rangle \left\langle g_{i}\right|\doteq\diag(\underbrace{0,0,\dots0}_{n\text{-zeros}},x_{g_{1}},x_{g_{2}}\dots x_{g_{n}}),
\]
\[
\left|w\right\rangle :=\sum_{i=1}^{n}\sqrt{p_{h_{i}}}\left|h_{i}\right\rangle \doteq(\sqrt{p_{h_{1}}},\sqrt{p_{h_{2}}}\dots\sqrt{p_{h_{n}}},\underbrace{0,0\dots0}_{n\text{-zeros}})^{T}
\]

\[
\left|v\right\rangle :=\sum_{i=1}^{n}\sqrt{p_{g_{i}}}\left|g_{i}\right\rangle \doteq(\underbrace{0,0\dots0}_{n\text{-zeros}},\sqrt{p_{g_{1}}},\sqrt{p_{g_{2}}}\dots\sqrt{p_{g_{n}}})^{T}.
\]
Then, 
$$
O:=\sum_{i=0}^{n-1}\left(\frac{\Pi_{h_{i-1}}^{\perp}(X_{h})^{i}\left|w\right\rangle \left\langle v\right|(X_{g})^{i}\Pi_{g_{i-1}}^{\perp}}{\sqrt{c_{h_{i}}c_{g_{i}}}}+\hc\right)
$$
satisfies 
\[
X_{h}\ge E_{h}OX_{g}O^{T}E_{h}\quad\text{and}\quad O\left|v\right\rangle =\left|w\right\rangle, 
\]
where $\Pi_{h_{-1}}^{\perp}=\Pi_{g_{-1}}^{\perp}=\mathbb{I}$, 
$
\Pi_{h_{i}}^{\perp}:=\text{projector orthogonal to }\text{span}\{(X_{h})^{i}\left|w\right\rangle ,(X_{h})^{i-1}\left|w\right\rangle ,\dots\left|w\right\rangle \},
$ \\ $c_{h_{i}}:=\left\langle w\right|(X_{h})^{i}\Pi_{h_{i-1}}^{\perp}(X_{h})^{i}\left|w\right\rangle $, 
and analogously are defined the forms of $\Pi_{g_{i}}^{\perp}$ and $c_{g_{i}}$.
\end{prop}

\begin{proof}
Let $t=\sum_{i=1}^{2n} p_i \llbracket x_i \rrbracket$ be the zeroth assignment.  \Lemref{expectationLemma} from Appendix \ref{app:lemmas} gives us the following properties of $t$:
\begin{eqnarray}
\left\langle x^{k}\right\rangle&=0,\ \ \ \ \ \  \text{ for all } k\in\{0,1,2\dots,2n-2\},\text{ and} \label{eq:Mochonsf0equality}\\ 
&\left\langle x^{2n-1}\right\rangle>0\label{eq:Mochonsf0Positivity}
\end{eqnarray}
where $\langle x^{k} \rangle := \sum_{i=1}^{2n} p_i (x_i)^k$.
Consider the following basis:
\begin{align}
\left|w_{0}\right\rangle  & :=\left|w\right\rangle \nonumber \\
\left|w_{1}\right\rangle  & :=\frac{\left(\mathbb{I}-\left|w_{0}\right\rangle \left\langle w_{0}\right|\right)(X_{h})\left|w\right\rangle }{\sqrt{c_{h_{1}}}}\nonumber \\
\vdots\nonumber \\
\left|w_{k}\right\rangle  & :=\frac{\left(\mathbb{I}-\sum_{i=0}^{k-1}\left|w_{i}\right\rangle \left\langle w_{i}\right|\right)(X_{h})^{k}\left|w\right\rangle }{\sqrt{c_{h_{k}}}}.\label{eq:w_k}
\end{align}
We are interested in keeping track of the highest power, $l$, of $\langle x_h^l \rangle $. To this end, we consider the highest power of $X_h$ that appears in $\ket{w_k}$, i.e. $X_h^k$ and the highest value $l'$ such that a $\langle x_h^{l'} \rangle$ appears in $\ket{w_k}$, i.e. $l'=2k$ (as $\langle x_h^{2k} \rangle$ is present in $\sqrt{c_{h_{k}}}$). We capture this dependence by writing $
\mathcal{M}(\left|w_{k}\right\rangle )=\left\langle x_{h}^{2k}\right\rangle \cdot(X_{h})^{k}\left|w\right\rangle  
$. %
Note that the projectors can
be expressed in terms of these vectors more concisely, as
$
\Pi_{h_{i}}:=\mathbb{I}-\Pi_{h_{i}}^{\perp}=\sum_{j=0}^{i}\left|w_{j}\right\rangle \left\langle w_{j}\right|.
$
It also follows that $O$ can be re-written as 
\[
O=\sum_{j=0}^{n-1}\left(\left|w_{j}\right\rangle \left\langle v_{j}\right|+\left|v_{j}\right\rangle \left\langle w_{j}\right|\right),
\]
where $\left|v_{j}\right\rangle $ is analogously defined (by replacing
$h$ with $g$). It is evident that $O\left|v\right\rangle =\left|w\right\rangle $.
We set $D=X_{h}-E_{h}OX_{g}O^{T}E_{h}$, and note that $\left\langle v_{j}\right|D\left|v_{i}\right\rangle =0$
(because $X_{h}\left|v_{i}\right\rangle =0$ and $E_{h}\left|v_{i}\right\rangle =0$\footnote{The conclusion holds even without the projector as $O$ maps $\text{span}(\left|v_{1}\right\rangle ,\left|v_{2}\right\rangle ,\dots\left|v_{n}\right\rangle )$
to $\text{span}(\left|w_{1}\right\rangle ,\left|w_{2}\right\rangle \dots\left|w_{n}\right\rangle )$
on which $X_{g}$ has no support.}). We assert that it has the following rank-1 form 
\[
D=\left[\begin{array}{cccc}
0 & \dots &  & 0\\
\vdots & \ddots &  & \vdots\\
0 & \dots &   & \left\langle w_{n-1}\right|D\left|w_{n-1}\right\rangle 
\end{array}\right]
\]
in the $\left(\left|w_{0}\right\rangle ,\left|w_{1}\right\rangle ,\dots\left|w_{n-1}\right\rangle \right)$
basis, together with $\left\langle w_{n-1}\right|D\left|w_{n-1}\right\rangle >0$.
To see this, we simply compute 
\begin{equation*}
\left\langle w_{i}\right|D\left|w_{j}\right\rangle  =\left\langle w_{i}\right|X_{h}\left|w_{j}\right\rangle -\left\langle w_{i}\right|OX_{g}O^{T}\left|w_{j}\right\rangle 
 =\left\langle w_{i}\right|X_{h}\left|w_{j}\right\rangle -\left\langle v_{i}\right|X_{g}\left|v_{j}\right\rangle .
\end{equation*}
For any $0\le i,j\le n-1$ except for the case where
both $i=j=n-1$, the two terms are the same. This is because the term
with the highest possible power $l$ (of $\left\langle x^{l}\right\rangle $)
in $\left\langle w_{i}\right|X_{h}\left|w_{j}\right\rangle $ can
be deduced by observing
\begin{equation}
\mathcal{M}(\left\langle w_{i}\right|)X_{h}\mathcal{M}(\left|w_{j}\right\rangle )=\left\langle x_{h}^{2i}\right\rangle \cdot\left\langle x_{h}^{2j}\right\rangle \cdot\left\langle x_{h}^{i+j+1}\right\rangle .\nonumber
\end{equation}
For the analogous expression with $g$ to be the same, we must have
$2i,2j$ and $i+j+1$ less than or equal to $2n-2$. The first two
are always satisfied (for $0\le i,j\le n-1$). The last can only be
violated when $i=j=n-1$. This establishes that the matrix has the
asserted form.

To prove the positivity of $\left\langle w_{n-1}\right|D\left|w_{n-1}\right\rangle $,
consider $\left\langle w_{n-1}\right|X_{h}\left|w_{n-1}\right\rangle $
and $\left\langle v_{n-1}\right|X_{g}\left|v_{n-1}\right\rangle $.
When these terms are expanded in powers of $\left\langle x_{h}^{k}\right\rangle $
and $\left\langle x_{g}^{k}\right\rangle $ respectively, only terms
with $k>2n-2$ would remain; the others would get cancelled due to
\Eqref{Mochonsf0equality}. Using \Eqref{w_k}, it follows that 
\[
\left\langle w_{n-1}\right|D\left|w_{n-1}\right\rangle =\frac{1}{c_{h_{n-1}}}\left\langle w\right|(X_{h})^{2n-2+1}\left|w\right\rangle -\frac{1}{c_{g_{n-1}}}\left\langle v\right|(X_{g})^{2n-2+1}\left|v\right\rangle,
\]
and it is not hard to see that $c_{h_{n-1}}=c_{h_{n-1}}(\left\langle x_{h}^{2n-2}\right\rangle ,\left\langle x_{h}^{2n-3}\right\rangle ,\dots,\left\langle x_{h}^{1}\right\rangle )$
does not depend on $\left\langle x_{h}^{2n-1}\right\rangle $ (we
proceed analogously for $c_{g_{n-1}}$). Further, $c_{h_{n-1}}=c_{g_{n-1}}=:c_{n-1}$.
We thus have 
\[
\left\langle w_{n-1}\right|D\left|w_{n-1}\right\rangle =\frac{\left\langle x_{h}^{2n-1}\right\rangle }{c_{n-1}}>0
\]
using \Eqref{Mochonsf0Positivity}. Thus, $X_{h}-E_h OX_{g}O^{T} E_h \ge0$. Note that we assumed $\text{span}\{\left|w\right\rangle ,X_{h}\left|w\right\rangle ,X_{h}^{2}\left|w\right\rangle ,\dots,X_{h}^{n}\left|w\right\rangle \}$
equals to $\text{span}\{\left|h_{1}\right\rangle ,\left|h_{2}\right\rangle \dots\left|h_{n}\right\rangle \}$
which is justified by \Lemref{spanningLemma}, presented in Appendix \ref{app:lemmas}.
\end{proof}

Before proceeding to the unbalanced zeroth assignments,
let us try to better understand the above result 
and see why it doesn't work unchanged in the unbalanced case. We could write $D_{ij}=\left\langle w_{i}\right|D\left|w_{j}\right\rangle $
and note that the maximum power $l$ which appears as $\left\langle x_{g/h}^{l}\right\rangle $
is given by $\max\{2i,2j,i+j+1\}$. This yields a matrix with each
term depending on the power as 
\[
D=\left[\begin{array}{cccccc}
D_{00}(\left\langle x\right\rangle )\\
D_{10}(\left\langle x^{2}\right\rangle ,\dots) & D_{11}(\left\langle x^{3}\right\rangle ,\dots) &  &  & \text{h.c.}\\
D_{20}(\left\langle x^{4}\right\rangle ,\dots) & D_{21}(\left\langle x^{4}\right\rangle ,\dots) & D_{22}(\left\langle x^{5}\right\rangle ,\dots)\\
D_{30}(\left\langle x^{6}\right\rangle ,\dots) & D_{31}(\left\langle x^{6}\right\rangle ,\dots) & D_{32}(\left\langle x^{6}\right\rangle ,\dots) & D_{33}(\left\langle x^{7}\right\rangle ,\dots)\\
D_{40}(\left\langle x^{8}\right\rangle ,\dots) & D_{41}(\left\langle x^{8}\right\rangle ,\dots) & D_{42}(\left\langle x^{8}\right\rangle ,\dots) & D_{43}(\left\langle x^{8}\right\rangle ,\dots) & D_{44}(\left\langle x^{9}\right\rangle ,\dots)\\
 &  &  &  &  & \ddots
\end{array}\right].
\]
For brevity, we represent this dependence as 
\[
\mathcal{M}(D)=\left[\begin{array}{cccc}
\left\langle x\right\rangle & & \text{h.c.} \\
\left\langle x^{2}\right\rangle  & \left\langle x^{3}\right\rangle \\
\left\langle x^{4}\right\rangle  & \left\langle x^{4}\right\rangle  & \left\langle x^{5}\right\rangle \\
 &  &  & \ddots
\end{array}\right].
\]
Consider the balanced $m_0$ case over $\{x_{1},x_{2},x_{3},x_{4}\}$, where we have
 $\left\langle x\right\rangle =\left\langle x^{2}\right\rangle =0$
and $\left\langle x^{3}\right\rangle >0$. This is a two-dimensional
case, thus 
\[
\mathcal{M}(D)=\left[\begin{array}{cc}
0 & 0\\
0 & \left\langle x^{3}\right\rangle 
\end{array}\right]\ge0.
\]
If we now try to use the same procedure for an unbalanced zeroth assignment
over $\left\{ x_{1},x_{2}\dots x_{5}\right\} $, we will have $\left\langle x\right\rangle =\left\langle x^{2}\right\rangle =\left\langle x^{3}\right\rangle =0$
and $\left\langle x^{4}\right\rangle >0$. If we try to solve in three
dimensions, we would obtain 
\begin{equation}
\mathcal{M}(D)=\left[\begin{array}{ccc}
0 & 0 & \left\langle x^{4}\right\rangle \\
0 & 0 & \left\langle x^{4}\right\rangle \\
\left\langle x^{4}\right\rangle  & \left\langle x^{4}\right\rangle  & \left\langle x^{5}\right\rangle 
\end{array}\right]\label{eq:exampleSubMatrix}
\end{equation}
which does not seem to work directly. It turns out that the projector
that was present in \Eqref{TEF_const}, gets rid of the troublesome
part and yields a zero matrix. We see it in this example first and
then generalize it. The unbalanced assignment takes three points to
two points. We define $X_{h}:=\diag(x_{h_{1}},x_{h_{2}},0,0,0)$,
$\left|w\right\rangle =(\sqrt{p_{h_{1}}},\sqrt{p_{h_{2}}},0,0,0)$
along with $\left|w_{0}\right\rangle:=\left|w\right\rangle \text{ and }
\left|w_{1}\right\rangle:=\left(\mathbb{I}-\left|w_{0}\right\rangle \left\langle w_{0}\right|\right)X_{h}\left|w_{0}\right\rangle .
$
We can write $E_{h}=\sum_{i=0}^{1}\left|w_{i}\right\rangle \left\langle w_{i}\right|$
and have the same orthogonal matrix as before, except that we leave $\left|v_{2}\right\rangle $
unchanged, i.e. $O=\sum_{i=0}^{1}\left|w_{i}\right\rangle \left\langle v_{i}\right|+\left|v_{2}\right\rangle \left\langle v_{2}\right|$.
We can now show that $D'=X_{h}-E_{h}OX_{g}O^{T}E_{h}\ge0$ because
every vector in $\left|\psi\right\rangle \in\text{span}\{\left|v_{0}\right\rangle ,\left|v_{1}\right\rangle ,\left|v_{2}\right\rangle \}$
satisfies $D'\left|\psi\right\rangle =0$ (as $X_{h}\left|\psi\right\rangle =0$
and $E_{h}\left|\psi\right\rangle =0$). This means that it suffices
to restrict to a $2\times2$ matrix in $\text{span}\{\left|w_{0}\right\rangle ,\left|w_{1}\right\rangle \}$.
But, from \Eqref{exampleSubMatrix}, we already know that this is zero,
hence $D'=0$.

\begin{prop}[Solution to unbalanced zeroth assignments]
Let $t=\sum_{i=1}^{n-1}p_{h_{i}}\llbracket x_{h_{i}}\rrbracket -\sum_{i=1}^{n}p_{g_{i}}\llbracket x_{g_{i}}\rrbracket$ be a zeroth assignment over $0<x_{1}<x_{2}\dots<x_{2n-1}$, $\left\{ \left|h_{1}\right\rangle ,\left|h_{2}\right\rangle \dots\left|h_{n-1}\right\rangle ,\left|g_{1}\right\rangle ,\left|g_{2}\right\rangle \dots\left|g_{n}\right\rangle \right\} $
be an orthonormal basis, $E_{h}:=\sum_{i=1}^{n}\left|h_{i}\right\rangle \left\langle h_{i}\right|$ a subspace projector,
and finally let

\[
X_{h}:=\sum_{i=1}^{n-1}x_{h_{i}}\left|h_{i}\right\rangle \left\langle h_{i}\right|\doteq\diag(x_{h_{1}},x_{h_{2}}\dots x_{h_{n-1}},\underbrace{0,0,\dots0}_{n\text{ zeros}}),
\]
\[
X_{g}:=\sum_{i=1}^{n}x_{g_{i}}\left|g_{i}\right\rangle \left\langle g_{i}\right|\doteq\diag(\underbrace{0,0,\dots0}_{n-1\text{ zeros}},x_{g_{1}},x_{g_{2}}\dots x_{g_{n-1}},x_{g_{n}}),
\]
 
\[
\left|w\right\rangle :=\sum_{i=1}^{n-1}\sqrt{p_{h_{i}}}\left|h_{i}\right\rangle \doteq(\sqrt{p_{h_{1}}},\sqrt{p_{h_{2}}},\dots\sqrt{p_{h_{n-1}}},\underbrace{0,0\dots0}_{n\text{ zeros}})^{T},
\]
\[
\left|v\right\rangle :=\sum_{i=1}^{n}\sqrt{p_{g_{i}}}\left|g_{i}\right\rangle \doteq(\underbrace{0,0,\dots0}_{n-1\text{ zeros}},\sqrt{p_{g_{1}}},\sqrt{p_{g_{2}}}\dots\sqrt{p_{g_{n-1}}},\sqrt{p_{g_{n}}})^{T}.
\]
 
$$\text{Then, }\ \ \ \ \ \ \ \ \ \ \ \ \  \ \ \ \ \ \ \ \  \ \ \ \ \ \ \ \ 
O:=\left(\sum_{i=0}^{n-2}\frac{\Pi_{h_{i-1}}^{\perp}(X_{h})^{i}\left|w\right\rangle \left\langle v\right|(X_{g})^{i}\Pi_{g_{i-1}}^{\perp}}{\sqrt{c_{h_{i}}c_{g_{i}}}}+\hc\right)+\frac{\Pi_{g_{n-2}}^{\perp}(X_{g})^{n-1}\left|v\right\rangle \left\langle v\right|(X_{g})^{n-1}\Pi_{g_{n-2}}^{\perp}}{c_{g_{i}}}
$$
satisfies 
\[
X_{h}\ge E_{h}OX_{g}O^{T}E_{h}\quad\text{and}\quad E_{h}O\left|v\right\rangle =\left|w\right\rangle,
\]
where $\Pi_{h_{-1}}^{\perp}=\Pi_{g_{-1}}^{\perp}=\mathbb{I}, \ \Pi_{h_{i}}^{\perp}:=\text{projector orthogonal to }\text{span}\{(X_{h})^{i}\left|w\right\rangle ,(X_{h})^{i-1}\left|w\right\rangle ,\dots\left|w\right\rangle \},$\\
$c_{h_{i}}:=\left\langle w\right|(X_{h})^{i}\Pi_{h_{i-1}}^{\perp}(X_{h})^{i}\left|w\right\rangle $, 
and analogous are the forms of $\Pi_{g_{i}}^{\perp}$ and $c_{g_{i}}$.
\label{prop:mochonf0unbalanced}
\end{prop}
\begin{proof}
By using again \Lemref{expectationLemma} from Appendix \ref{app:lemmas}, we have
\begin{equation}
\left\langle x^{k}\right\rangle =0\ \ \text{for}\ \ \  k\in\{0,1,\dots2n-3\},
\label{eq:Mochonf0equalityunbal}
\end{equation}
and 
\begin{equation}
    \left\langle x^{2n-2}\right\rangle >0.\nonumber
\end{equation}

We define the basis, almost exactly as before, we set $\left|w_{0}\right\rangle :=\left|w\right\rangle $
and for each integer $k$ satisfying $0\le k\le n-2$ we have
\[
\left|w_{k}\right\rangle :=\frac{\Pi_{h_{k-1}}^{\perp}(X_{h})^{k}\left|w\right\rangle }{\sqrt{c_{h_{k}}}}=\frac{\left(\mathbb{I}-\sum_{i=0}^{k-1}\left|w_{i}\right\rangle \left\langle w_{i}\right|\right)(X_{h})^{k}\left|w\right\rangle }{\sqrt{c_{h_{k}}}}.
\]
We define $\left|v_{0}\right\rangle :=\left|v\right\rangle $ and
for each integer satisfying $0\le k\le n-1$ we have 
\[
\left|v_{k}\right\rangle :=\frac{\Pi_{g_{k-1}}^{\perp}(X_{g})^{k}\left|v\right\rangle }{\sqrt{c_{g_{k}}}}=\frac{\left(\mathbb{I}-\sum_{i=0}^{k-1}\left|v_{i}\right\rangle \left\langle v_{i}\right|\right)(X_{g})^{k}\left|v\right\rangle }{\sqrt{c_{h_{k}}}}.
\]
Note that this means $O=\sum_{i=0}^{n-2}\left(\left|w_{i}\right\rangle \left\langle v_{i}\right|+\left|v_{i}\right\rangle \left\langle w_{i}\right|\right)+\left|v_{n-1}\right\rangle \left\langle v_{n-1}\right|$
and so $E_{h}O\left|v\right\rangle =\left|w\right\rangle $ follows
directly. Also, to establish $D:=X_{h}-E_{h}OX_{g}O^{T}E_{h}\ge0$,
note that it suffices to show that $\left\langle w_{i}\right|D\left|w_{j}\right\rangle \ge0$
for integers $i,j$ satisfying $0\le i,j\le n-2$. This is because,
as we saw in the previous case, $D\left|v_{i}\right\rangle =0$ as
$X_{h}\left|v_{i}\right\rangle =0$ and $E_{h}\left|v_{i}\right\rangle =0$.
As before, we indicate the term with the highest power of $X_{h}$
appearing in $\left|w_{k}\right\rangle $, for $k$ in $\{0,1\dots n-2\}$,
by 
\[
\mathcal{M}(\left|w_{k}\right\rangle )=\left\langle x_{h}^{2k}\right\rangle \cdot(X_{h})^{k}\left|w\right\rangle 
\]
and analogously, the highest power of $X_{g}$ appearing in $\left|v_{k}\right\rangle $
for $k$ in $\{0,1,\dots n-2\}$, by 
\[
\mathcal{M}(\left|v_{k}\right\rangle )=\left\langle x_{g}^{2k}\right\rangle \cdot(X_{g})^{k}\left|v\right\rangle .
\]
Again, the highest power $l$ of $\left\langle x^{l}\right\rangle $
that appears in $\left\langle w_{i}\right|D\left|w_{j}\right\rangle $
is $\max\{2j,2i,i+j+1\}$ which can be deduced by evaluating 
\[
\mathcal{M}(\left\langle w_{i}\right|)X_{h}\mathcal{M}(\left|w_{j}\right\rangle )=\left\langle x_{h}^{2j}\right\rangle \cdot\left\langle x_{h}^{2i}\right\rangle \cdot\left\langle x_{h}^{i+j+1}\right\rangle 
\]
and similarly 
\[
\mathcal{M}(\left\langle v_{i}\right|)E_{h}OX_{g}OE_{h}\mathcal{M}(\left|v_{i}\right\rangle )=\left\langle x_{g}^{2j}\right\rangle \cdot\left\langle x_{g}^{2i}\right\rangle \cdot\left\langle x_{g}^{i+j+1}\right\rangle .
\]
The highest possible power is obtained when $i=j=n-2$. This yields
$2n-3$ and thus, using \Eqref{Mochonf0equalityunbal}, we conclude that
$\left\langle w_{i}\right|D\left|w_{j}\right\rangle $ is zero for
all $0\le i,j\le n-2$, establishing in fact that $D=0$.
\end{proof}

\section{Solution to the monomial assignments  }

In this section we present the solutions to the monomial assignments of order higher than zero. 
There are four different cases, depending on the number of points and the degree of the monomial (balanced/unbalanced and aligned/misaligned, see \Defref{f_assignment-f_0_assignment-balanced-m_kmonomial-aligned}). One could find a single expression
for all, but this does not seem to aid clarity, therefore we present and prove the four cases separately. Our approach is essentially the same as before. The main additional technique that we introduce here is the use of the pseudo-inverses $X_{h}^{\dashv}$ and $X_{g}^{\dashv}$.\footnote{For any Hermitian matrix $A$ with spectral decomposition $A=\sum_i a_i\ket{i}\bra{i}$ (including zero eigenvalues), we denote by $A^{\dashv}$ its pseudo-inverse $A^{\dashv}:=\sum_{i:|a_i|>0}a_i^{-1}\ket{i}\bra{i}$. }
\begin{prop}[Solution to balanced aligned monomial assignments]
 \label{prop:ExactSolnBalancedMonomialAligned}Let $m=2b$ be an even non-negative integer, $
t=\sum_{i=1}^{n}x_{h_{i}}^{m}p_{h_{i}}\llbracket x_{h_{i}}\rrbracket -\sum_{i=1}^{n}x_{g_{i}}^{m}p_{g_{i}}\llbracket x_{g_{i}}\rrbracket
$ a monomial assignment over $0<x_{1}<x_{2}\dots<x_{2n}$, $\left\{ \left|h_{1}\right\rangle ,\left|h_{2}\right\rangle \dots\left|h_{n}\right\rangle ,\left|g_{1}\right\rangle ,\left|g_{2}\right\rangle \dots\left|g_{n}\right\rangle \right\} $ an orthonormal basis, and finally let 
\[
X_{h}:=\sum_{i=1}^{n}x_{h_{i}}\left|h_{i}\right\rangle \left\langle h_{i}\right|\doteq\diag(x_{h_{1}},x_{h_{2}}\dots x_{h_{n}},\underbrace{0,0\dots0}_{n\text{ zeros}}),
\]
\[
X_{g}:=\sum_{i=1}^{n}x_{g_{i}}\left|g_{i}\right\rangle \left\langle g_{i}\right|\doteq\diag(\underbrace{0,0,\dots0}_{n\text{ zeros}},x_{g_{1}},x_{g_{2}}\dots x_{g_{n}}),
\]
\[
\left|w\right\rangle :=\sum_{i=1}^{n}\sqrt{p_{h_{i}}}\left|h_{i}\right\rangle \doteq(\sqrt{p_{h_{1}}},\sqrt{p_{h_{2}}}\dots\sqrt{p_{h_{n}}},\underbrace{0,0,\dots0}_{n\text{ zeros}})^{T}\ \ \ \text{and}\ \ \ \left|w'\right\rangle :=(X_{h})^{b}\left|w\right\rangle,
\]
\[
\left|v\right\rangle :=\sum_{i=1}^{n}\sqrt{p_{g_{i}}}\left|g_{i}\right\rangle \doteq(\underbrace{0,0,\dots0}_{n\text{ zeros}},\sqrt{p_{g_{1}}},\sqrt{p_{g_{2}}}\dots\sqrt{p_{g_{n}}})^{T}\ \ \ \text{and}\ \ \ \left|v'\right\rangle :=(X_{g})^{b}\left|v\right\rangle.
\] 

Then,
\[
O:=\sum_{i=-b}^{n-b-1}\left(\frac{\Pi_{h_{i}}^{\perp}(X_{h})^{i}\left|w'\right\rangle \left\langle v'\right|(X_{g})^{i}\Pi_{g_{i}}^{\perp}}{\sqrt{c_{h_{i}}c_{g_{i}}}}+\hc\right)
\]
 
satisfies
\[ 
X_{h}\ge E_{h}OX_{g}O^{T}E_{h}\quad and\quad E_{h}O\left|v'\right\rangle =\left|w'\right\rangle,
\]

where $E_{h}:=\sum_{i=1}^{n}\left|h_{i}\right\rangle \left\langle h_{i}\right|$, and for brevity, by $X_{h}^{-k}$ we mean $(X_{h}^{\dashv})^{k}$
for $k>0$ (similarly for $X_{g}$), 

\[
\Pi_{h_{i}}^{\perp}:=\begin{cases}
\text{projector orthogonal to }\text{span}\{(X_{h})^{-|i|+1}\left|w'\right\rangle ,(X_{h})^{-|i|+2}\left|w'\right\rangle \dots,\left|w'\right\rangle \} & i<0\\
\text{projector orthogonal to }\text{span}\{(X_{h})^{-b}\left|w'\right\rangle ,(X_{h})^{-b+1}\left|w'\right\rangle ,\dots(X_{h})^{i-1}\left|w'\right\rangle \} & i>0\\
\mathbb{I} & i=0,
\end{cases}
\]
$c_{h_{i}}:=\left\langle w'\right|(X_{h})^{i}\Pi_{h_{i}}^{\perp}(X_{h})^{i}\left|w'\right\rangle $, and analogous are the forms of $\Pi_{g_{i}}^{\perp}$ and $c_{g_{i}}$. 
\end{prop}

\begin{prop}[Solution to balanced misaligned monomial assignments]
 \label{prop:ExactSolnBalancedMonomialMisaligned}Let $m=2b-1$ be an odd non-negative integer, $
t=\sum_{i=1}^{n}x_{h_{i}}^{m}p_{h_{i}}\llbracket x_{h_{i}}\rrbracket -\sum_{i=1}^{n}x_{g_{i}}^{m}p_{g_{i}}\llbracket x_{g_{i}}\rrbracket ,
$ a monomial assignment over $0<x_{1}<x_{2}\dots<x_{2n}$, $\left\{ \left|h_{1}\right\rangle ,\left|h_{2}\right\rangle \dots\left|h_{n}\right\rangle ,\left|g_{1}\right\rangle ,\left|g_{2}\right\rangle \dots\left|g_{n}\right\rangle \right\} $ an orthonormal basis, and finally let
\[
X_{h}:=\sum_{i=1}^{n}x_{h_{i}}\left|h_{i}\right\rangle \left\langle h_{i}\right|\doteq\diag(x_{h_{1}},x_{h_{2}}\dots x_{h_{n}},\underbrace{0,0\dots0}_{n\text{ zeros}}),
\]
\[
X_{g}:=\sum_{i=1}^{n}x_{g_{i}}\left|g_{i}\right\rangle \left\langle g_{i}\right|\doteq\diag(\underbrace{0,0,\dots0}_{n\text{ zeros}},x_{g_{1}},x_{g_{2}}\dots x_{g_{n}}),
\]
\[
\left|w\right\rangle :=(\sqrt{p_{h_{1}}},\sqrt{p_{h_{2}}}\dots\sqrt{p_{h_{n}}},\underbrace{0,0\dots0}_{n\text{ zeros}})\ \ \ \text{and}\ \ \ \left|w'\right\rangle :=(X_{h})^{b-\frac{1}{2}}\left|w\right\rangle,
\]
\[
\left|v\right\rangle :=(\underbrace{0,0,\dots0}_{n\text{ zeros}},\sqrt{p_{g_{1}}},\sqrt{p_{g_{2}}}\dots\sqrt{p_{g_{n}}})\ \ \ \text{and}\ \ \ \left|v'\right\rangle :=(X_{g})^{b-\frac{1}{2}}\left|v\right\rangle.
\]
 
\begin{align*}
\text{Then, }\ \ \ \ \ \ \ \ \ O & :=\sum_{i=-b+1}^{n-b-1}\left(\frac{\Pi_{h_{i}}^{\perp}(X_{h})^{i}\left|w'\right\rangle \left\langle v'\right|(X_{g})^{i}\Pi_{g_{i}}^{\perp}}{\sqrt{c_{h_{i}}c_{g_{i}}}}+\hc\right)\\
 & \quad+\frac{\Pi_{g_{n-b}}^{\perp}(X_{g})^{n-b}\left|v'\right\rangle \left\langle v'\right|(X_{g})^{n-b}\Pi_{g_{n-b}}^{\perp}}{c_{g_{n-b+1}}}+\frac{\Pi_{h_{n-b}}^{\perp}(X_{h})^{n-b}\left|w'\right\rangle \left\langle w'\right|(X_{h})^{n-b}\Pi_{h_{n-b}}^{\perp}}{c_{h_{n-b}}}
\end{align*}

satisfies
\[
X_{h}\ge E_{h}OX_{g}O^{T}E_{h}\quad and\quad E_{h}O\left|v'\right\rangle =\left|w'\right\rangle, 
\]
where $E_{h}:=\sum_{i=1}^{n}\left|h_{i}\right\rangle \left\langle h_{i}\right|$, and for brevity, by $X_{h}^{-k}$ we mean $(X_{h}^{\dashv})^{k}$
for $k>0$ (similarly for $X_{g}$),
\[
\Pi_{h_{i}}^{\perp}:=\begin{cases}
\text{projector orthogonal to }\text{span}\{(X_{h}^{\dashv})^{|i|-1}\left|w'\right\rangle ,(X_{h}^{\dashv})^{|i|-2}\left|w'\right\rangle \dots,\left|w'\right\rangle \} & i<0\\
\text{projector orthogonal to }\text{span}\{(X_{h}^{\dashv})^{b-1}\left|w'\right\rangle ,(X_{h}^{\dashv})^{b-2}\left|w'\right\rangle ,\dots,\left|w'\right\rangle ,X_{h}\left|w'\right\rangle ,\dots(X_{h})^{i-1}\left|w'\right\rangle \} & i>0\\
\mathbb{I} & i=0,
\end{cases}
\]
 $c_{h_{i}}:=\left\langle w'\right|(X_{h})^{i}\Pi_{h_{i}}^{\perp}(X_{h})^{i}\left|w'\right\rangle $, and analogous are the forms of $\Pi_{g_{i}}^{\perp}$ and $c_{g_i}$.

\end{prop}

For the proofs and concrete examples of balanced aligned and misaligned monomial assignments, see Appendix \ref{app:BalancedMon}.

We similarly proceed to the unbalanced monomial assignments, aligned and misaligned. Below, we state the solution for both cases, while in Appendix \ref{app:UnbalancedMon} we prove their correctness and give concrete examples illustrating their construction. 

\begin{prop}[Solution to the unbalanced aligned monomial assignments]
\label{prop:ExactSolnUnbalancedMonomialAligned} Let $m=2b$ be an even non-negative integer, $
t=\sum_{i=1}^{n-1}x_{h_{i}}^{m}p_{h_{i}}\llbracket x_{h_{i}}\rrbracket -\sum_{i=1}^{n}x_{g_{i}}^{m}p_{g_{i}}\llbracket x_{g_{i}}\rrbracket$ a monomial assignment over $0<x_{1}<x_{2}\dots<x_{2n-1}$,  $\left\{ \left|h_{1}\right\rangle ,\left|h_{2}\right\rangle \dots\left|h_{n-1}\right\rangle ,\left|g_{1}\right\rangle ,\left|g_{2}\right\rangle \dots\left|g_{n}\right\rangle \right\} $
be an orthonormal basis, and finally let
\[
X_{h}:=\sum_{i=1}^{n-1}x_{h_{i}}\left|h_{i}\right\rangle \left\langle h_{i}\right|\doteq\diag(x_{h_{1}},x_{h_{2}}\dots x_{h_{n-1}},\underbrace{0,0\dots0}_{n\text{ zeros}}),
\]
\[
X_{g}:=\sum_{i=1}^{n}x_{g_{i}}\left|g_{i}\right\rangle \left\langle g_{i}\right|\doteq\diag(\underbrace{0,0,\dots0}_{n-1\text{ zeros}},x_{g_{1}},x_{g_{2}}\dots x_{g_{n}}),
\]
\[
\left|w\right\rangle :=(\sqrt{p_{h_{1}}},\sqrt{p_{h_{2}}}\dots\sqrt{p_{h_{n-1}}},\underbrace{0,0\dots0}_{n\text{ zeros}})\ \ \  \text{and}\ \ \ \left|w'\right\rangle :=(X_{h})^{b}\left|w\right\rangle,
\]
\[
\left|v\right\rangle :=(\underbrace{0,0,\dots0}_{n-1\text{ zeros}},\sqrt{p_{g_{1}}},\sqrt{p_{g_{2}}}\dots\sqrt{p_{g_{n}}})\ \ \  \text{and}\ \ \ \left|v'\right\rangle :=(X_{g})^{b}\left|v\right\rangle.
\]

\begin{equation*}
\text{Then, }\ \ \ O :=\sum_{i=-b}^{n-b-2}\left(\frac{\Pi_{h_{i}}^{\perp}(X_{h})^{i}\left|w'\right\rangle \left\langle v'\right|(X_{g})^{i}\Pi_{g_{i}}^{\perp}}{\sqrt{c_{h_{i}}c_{g_{i}}}}+\hc\right)+\frac{\Pi_{g_{n-b-1}}^{\perp}(X_{g})^{n-b-1}\left|v'\right\rangle \left\langle v'\right|(X_{g})^{n-b-1}\Pi_{g_{n-b-1}}^{\perp}}{c_{g_{n-b-1}}}
\end{equation*}
satisfies 
\[
X_{h}\ge E_{h}OX_{g}O^{T}E_{h}\quad and\quad E_{h}O\left|v'\right\rangle =\left|w'\right\rangle, 
\]
where for brevity, by $X_{h}^{-k}$ we mean $(X_{h}^{\dashv})^{k}$
for $k>0$ (similarly for $X_{g}$), $c_{h_{i}},c_{g_{i}},\Pi_{h_{i}}^{\perp},\Pi_{g_{i}}^{\perp}$
are as defined in \Propref{ExactSolnBalancedMonomialAligned}. 
\end{prop}

\begin{prop}[Solution to the unbalanced misaligned monomial assignments]
\label{prop:ExactSolnUnbalancedMonomialMisaligned}Let $m=2b-1$ be an odd non-negative integer,  $t=\sum_{i=1}^{n}x_{h_{i}}^{m}p_{h_{i}}\llbracket x_{h_{i}}\rrbracket -\sum_{i=1}^{n-1}x_{g_{i}}^{m}p_{g_{i}}\llbracket x_{g_{i}}\rrbracket$ a monomial assignment over $0<x_{1}<x_{2}\dots<x_{2n-1}$,  $\left\{ \left|h_{1}\right\rangle ,\left|h_{2}\right\rangle \dots\left|h_{n}\right\rangle ,\left|g_{1}\right\rangle ,\left|g_{2}\right\rangle \dots\left|g_{n-1}\right\rangle \right\} $
be an orthonormal basis, and finally let

\[
X_{h}:=\sum_{i=1}^{n}x_{h_{i}}\left|h_{i}\right\rangle \left\langle h_{i}\right|\doteq\diag(x_{h_{1}},x_{h_{2}}\dots x_{h_{n}},\underbrace{0,0\dots0}_{n-1\text{ zeros}}),
\]
\[
X_{g}:=\sum_{i=1}^{n-1}x_{g_{i}}\left|g_{i}\right\rangle \left\langle g_{i}\right|\doteq\diag(\underbrace{0,0,\dots0}_{n\text{ zeros}},x_{g_{1}},x_{g_{2}}\dots x_{g_{n-1}}),
\]
\[
\left|w\right\rangle :=(\sqrt{p_{h_{1}}},\sqrt{p_{h_{2}}}\dots\sqrt{p_{h_{n}}},\underbrace{0,0\dots0}_{n-1\text{ zeros}})\ \ \ \text{and}\ \ \ \left|w'\right\rangle :=(X_{h})^{b-\frac{1}{2}}\left|w\right\rangle,
\]
\[
\left|v\right\rangle :=(\underbrace{0,0,\dots0}_{n\text{ zeros}},\sqrt{p_{g_{1}}},\sqrt{p_{g_{2}}}\dots\sqrt{p_{g_{n-1}}})\ \ \ \text{and}\ \ \ \left|v'\right\rangle :=(X_{g})^{b-\frac{1}{2}}\left|v\right\rangle.
\]

\begin{equation*}
\text{Then, }\ \ \ \ \ \ \ \ \ \ \ \ \ \ \ \ \ \ \ \ \ O :=\sum_{i=-b+1}^{n-b-1}\left(\frac{\Pi_{h_{i}}^{\perp}(X_{h})^{i}\left|w'\right\rangle \left\langle v'\right|(X_{g})^{i}\Pi_{g_{i}}^{\perp}}{\sqrt{c_{h_{i}}c_{g_{i}}}}+\hc\right)
+\frac{\Pi_{h_{n-b}}^{\perp}(X_{h})^{n-b}\left|w'\right\rangle \left\langle w'\right|(X_{h})^{n-b}\Pi_{h_{n-b}}^{\perp}}{c_{h_{n-b}}}
\end{equation*}
satisfies 
\[
X_{h}\ge E_{h}OX_{g}O^{T}E_{h}\quad and\quad E_{h}O\left|v'\right\rangle =\left|w'\right\rangle, 
\]
where for brevity, by $X_{h}^{-k}$ we mean $(X_{h}^{\dashv})^{k}$
for $k>0$ (similarly for $X_{g}$), $c_{h_{i}},c_{g_{i}},\Pi_{h_{i}}^{\perp},\Pi_{g_{i}}^{\perp}$
are as defined in \Propref{ExactSolnBalancedMonomialMisaligned}.
\end{prop} 

Combining all the above, we can now state our main result:

\begin{thm}
Let $t$ be an $f$-assignment (see \Defref{f_assignment-f_0_assignment-balanced-m_kmonomial-aligned}) with $f$ having real positive roots.
Then, in order to obtain its effective solution (see \Defref{solvinassign}), it suffices to write it as $t=\sum_{i}\alpha_{i}t_{i}'$ (see \Lemref{generalMonomialDecomposition}),
where $\alpha_{i}$ are positive and $t_{i}'$ are monomial assignments.
Furthermore, each monomial assignment $t_{i}'$ admits an exact solution given in \Propref{ExactSolnBalancedMonomialAligned},
\Propref{ExactSolnBalancedMonomialMisaligned}, \Propref{ExactSolnUnbalancedMonomialAligned},
or \Propref{ExactSolnUnbalancedMonomialMisaligned}. \label{thm:Main}
\end{thm}

\begin{proof}
We established that in order to determine the effective solution to an $f-$assignment $t$,  it is sufficient to express it as a sum of monomial assignments $t_{i'}$ and find the solution for each one of them (see Appendix \ref{app:sum-of-valid-functions}).
A monomial assignment can be balanced/unbalanced and
aligned/misaligned (see \Defref{f_assignment-f_0_assignment-balanced-m_kmonomial-aligned}). The solution in each case is given by either \Propref{ExactSolnBalancedMonomialAligned},
\Propref{ExactSolnBalancedMonomialMisaligned}, \Propref{ExactSolnUnbalancedMonomialAligned},
or \Propref{ExactSolnUnbalancedMonomialMisaligned}.
\end{proof}

In Appendix \ref{app:example}, as an example, we describe how \Thmref{Main} can be applied to derive a WCF protocol with bias approaching $\frac{1}{14}$.

\section{Conclusions and future work}
We presented the analytical construction of explicit WCF protocols achieving arbitrarily close to zero bias, by means of Mochon's family of TDPGs \cite{Mochon07}, described by the respective $f-$assignments. Using the TEF from \cite{Arora2019}, these TDPGs can be converted into WCF protocols with the corresponding bias.
In order to obtain the solution for an $f-$assignment, we argued that it suffices to write it as a sum of  monomial assignments and find the solution for each term of the sum separately. For all four different types of monomial assignments, we constructed the corresponding solutions and proved that indeed satisfy the required conditions as stated in \Eqref{TEF_const} and the analysis following it. Importantly enough, our approach does not use the reduction of EBM functions to valid functions and it admits, thus, a simple and clear description. 
We also presented an example illustrating the construction of a WCF protocol with bias $\frac{1}{14}$.

There exist several related problems that deserve further study.
First, one could try to find analytic solutions corresponding to $f-$assignments in fewer
dimensions (assuming that they exist). This way, the only shortcoming of our approach  concerning resource requirements could be improved: while expressing the $f-$assignment as a sum of monomial assignments we are increasing
the dimensions, which in turn corresponds to an increase in the number
of qubits required. One could also try to find analytic solutions for the Pelchat-Hoyer point games \cite{pelchat13}, which is another family of point games giving rise to WCF protocols with arbitrarily close to zero bias. Moreover, given
the recently improved bound on the number of rounds of communication needed to achieve a certain bias $\epsilon$ \cite{Miller2019}, one can investigate whether there exist protocols matching these bounds. Finally, while one expects the bias to increase in the presence of noise, a thorough study of such
effects is needed in order to determine the robustness of WCF protocols against noise.
 
\section*{Acknowledgements}
We are thankful to Tom Van Himbeeck, Kishor Bharti, Stefano Pironio and Ognyan Oreshkov for various insightful discussions. We acknowledge support from the Belgian Fonds de la Recherche Scientifique -- FNRS under grant no R.50.05.18.F (QuantAlgo). The QuantAlgo project has received funding from the QuantERA ERA-NET Cofund in
Quantum Technologies implemented within the European Union's Horizon 2020 Programme. ASA further acknowledges the FNRS for support through the FRIA grants, 3/5/5 -- MCF/XH/FC -- 16754 and F
3/5/5 -- FRIA/FC -- 6700 FC 20759.

\appendix

\section{Decomposing TEF functions into sums of TEF functions}
\label{app:sum-of-valid-functions}
In this first part of the appendix we present how one can construct a WCF protocol with bias $\epsilon$, by decomposing the TEF functions (i.e., the functions that satisfy \Eqref{TEF_const} for some unitary matrix $O$\footnote{As already mentioned, restricting to real matrices is enough (see \cite{Arora2019}), therefore we assume that the matrices $O$ are orthogonal without loss of generality.}) of a so-called \emph{time-independent point game} (TIPG)\footnote{TIPGs are presented and studied in numerous previous works \cite{Mochon07,Aharon2014,Arora2019}.} with the same bias $\epsilon$ into a sum of TEF functions. 
This way, we establish our claim that, to convert Mochon's TIPGs (achieving vanishing bias) which rely non-trivially only on transitions defined by $f$-assignments, it is sufficient to find an effective solution thereof. In particular, it is sufficient to express an $f$-assignment as a sum of monomial assignments and find the solution to each one of them. 
In \Lemref{setequality}, we show that the set of TEF functions is the same as the set of valid functions, which in turn is the same as the closure of the set of EBM functions.\footnote{and the same holds for the closure of the set of EBRM functions, see \cite{Arora2018,Arora2019}.} Henceforth, for simplicity, we only use the term valid functions.
Our demonstration requires techniques and results from previous works \cite{Mochon07,Aharon2014,Arora2018,Arora2019}, which we do not present here in detail; we only refer to them and outline how they are used in our analysis.
We recall from \cite{Mochon07,ACG+14} the basic idea behind the conversion of a TIPG into a TDPG (see, for e.g., the proof of Theorem 5 in \cite{ACG+14}). 
 The primary hinderance is that for applying a valid function in a TDPG, the places where the function is negative must already have points with at least as much weight. This corresponds to finding a time dependent ordering of the valid functions which define a TIPG, however, in general, TIPGs do not admit such simple orderings. This difficulty is surpassed by introducing the so-called \emph{catalyst state}, which is a set of points with vanishing weights. They are a scaled-down compensation for the negative weights which arise. In their presence, an accordingly scaled-down version of the valid functions can be applied, repeatedly, until their cumulative effect is essentially the same as that of having applied the valid functions unaltered. The catalyst state, after this procedure, is effectively unchanged. The weight of the catalyst state costs us an increase in the bias. However, the weight can be made arbitrarily small, at the expense of extra rounds of communication. 
 Our case is not very different. Suppose that the valid functions used in the TIPG are decomposed into a sum of valid functions. Let us call these valid functions (present in the decomposition), \emph{constituent functions}. Then, we can convert the TIPG into a TDPG which only uses the constituent functions by essentially using the same technique. This is because the difficulty in constructing TDPGs using the constituent functions is of the same nature.
In particular, it is possible that the constituent functions are negative at various locations, but there are no points present there. We can again use a catalyst state, scale the constituent functions accordingly, and proceed thereafter as in the original proof \cite{Aharon2014}, to obtain the corresponding TDPG. The TEF from \cite{Arora2018,Arora2019} is then applied for this TDPG resulting in a WCF protocol approaching the same bias as the TIPG that we started with, in the limit of infinite rounds of communication.

\begin{lem}[TEF = Closure of EBM = valid]
The set of the TEF functions (as defined above), the set of valid functions (for the definition, see e.g. \cite{Mochon07,Aharon2014})  and the closure of the set of the EBM functions (for the definition see \Secref{Preliminaries}) are the same.
\label{lem:setequality}
\end{lem}
\begin{proof}[Proof outline]
We start by observing that the set of EBM functions is an open set. From \Defref{EBM_function} we can see that the matrix $H$ may have eigenvectors which have no support on $\ket{\psi}$. Consequently, one can consider a sequence of EBM functions ${t_i}$ such that the $\lim_{i\to\infty} t_i = t$ is well-defined, while the associated matrix $\lim_{i\to\infty} H_i$ has a diverging eigenvalue. Such a case arises, for instance, when we have a merge move in the point game. For concreteness, let $x_{g_1}, x_{g_2}$  be the coordinates of two points that are going to be merged into a single point with coordinate $x_h=p_{g_1}x_{g_1} + p_{g_2}x_{g_2}$, and let $p_{g_1}, p_{g_2}$ be their respective probability weights, with $p_{g_1}+p_{g_2}=1$. Furthermore, let $t_i = \llbracket x_h + 1/i \rrbracket - p_{g_1}\llbracket x_{g_1} \rrbracket - p_{g_2} \llbracket x_{g_2} \rrbracket$.  One can verify that for all finite values of $i$, $t_i$ is EBM, but its limit $t= \llbracket x_h \rrbracket - p_{g_1}\llbracket x_{g_1} \rrbracket - p_{g_2} \llbracket x_{g_2} \rrbracket$ is not EBM (we omit the details for the sake of brevity), thus concluding that the set of EBM functions is open.

To show that the closure of this set is the same as the set of the TEF functions, we need to establish that the limit of any such sequence belongs to the set of TEF functions. This requires a combination of certain results from Section 5 of \cite{Arora2018}. In particular, the relationship between the so-called \emph{canonical orthogonal form} and the \emph{canonical projective form} permits one to trade the divergence of such a matrix $H$ for appropriate projectors. This is exactly the origin of the projectors $E_h$ that appear in our analysis. 
The matrices $H \ge G$ and the vector $\ket{\psi}$ corresponding to an EBM transition, can be expressed in the canonical orthogonal form,\footnote{$X_h$ and $X_g$ are diagonal matrices containing the eigenvalues of $H$ and $G$, respectively. We suppress further details.} $X_h \ge O X_g O^T$. Essentially, the same orthogonal matrix $O$ also satisfies the TEF inequality.\footnote{The TEF inequality is closely related to the canonical projective form.} (\Eqref{TEF_const}) The TEF inequality may, in fact, be seen as the limit where $H$'s eigenvalues diverge to infinity. Thus, the limit $t$ of the sequence $t_i$ indeed belongs to the set of TEF functions and this argument readily extends to all relevant sequences.

Finally, in Section 3 of \cite{Aharon2014} the authors prove that the set of valid functions is the same as the closure of the set of EBM functions. In particular, they start by observing that the set of EBM functions is a convex cone $K$, and its dual cone $K^*$ is the set of operator monotone functions. The bi-dual $K^{**}$ is the set of valid functions, and the fact that  $K^{**}=\text{cl}(K)$ completes the proof. Since we just showed that the closure of the set of EBM functions is the same as the set of TEF functions, we can also conclude that the set of valid functions is the same as the set of TEF functions.

\end{proof}

\section{Useful lemmas}
\label{app:lemmas}

\begin{lem}
Consider a set of real coordinates $0\le x_{1}<x_{2}\dots<x_{n}$
and let $f(x)=(a_{1}-x)(a_{2}-x)\dots(a_{k}-x)$, where $k\le n-2$
and the roots $\{a_{i}\}_{i=1}^{k}$ of $f$ are non-negative. Let
$t=\sum_{i=1}^{n}p_{i}\left[ x_{i}\right] $ be
the corresponding $f$-assignment. Consider a set of real
coordinates $0<x_{1}+c<x_{2}+c\dots<x_{n}+c$, where $c>0$
and let $f'(x)=(a_{1}+c-x)(a_{2}+c-x)\dots(a_{k}+c-x)$. Let $t'=\sum_{i=1}^{n}p_{i}'\left[ x_{i}'\right] $
be the corresponding $f$-assignment with $x'_{i}:=x_{i}+c$.
The solution to $t$ and to $t'$ are the same. \label{lem:OriginIssueHandled}
\end{lem}
\begin{proof}
Note that $p'_{i}=p_{i}$ as the $c$'s cancel. We write $t=\sum_{i=1}^{n_{h}}p_{h_{i}}\left\llbracket x_{h_{i}}\right\rrbracket -\sum_{i=1}^{n_{g}}p_{g_{i}}\left\llbracket x_{g_{i}}\right\rrbracket $
and define $X_{h}:=\sum_{i=1}^{n_{h}}x_{h_{i}}\left|h_{i}\right\rangle $,
$X_{g}:=\sum_{i=1}^{n_{g}}x_{g_{i}}\left|g_{i}\right\rangle $. If
$t$ is solved by $O$, then we must have $X_{h}\ge E_{h}OX_{g}O^{T}E_{h}$.
We show that $X_{h}+c\mathbb{I}_{h}\ge E_{h}O(X_{g}+c\mathbb{I}_{g})O^{T}E_{h}$
where $\mathbb{I}_{h}:=\sum_{i=1}^{n_{h}}\left|h_{i}\right\rangle \left\langle h_{i}\right|$
and $\mathbb{I}_{g}:=\sum_{i=1}^{n_{g}}\left|g_{i}\right\rangle \left\langle g_{i}\right|$.
Together with the observation that $p'_{i}=p_{i}$, this establishes
that $O$ also solves $t'$. Since $c$ is an arbitrary real number,
it follows that $O$ solves $t$ if and only if it solves $t'$.

We now establish $X_{h}\ge E_{h}OX_{g}O^{T}E_{h}\iff X_{h}+c\mathbb{I}_{h}\ge E_{h}O(X_{g}+c\mathbb{I}_{g})O^{T}E_{h}$.
Observe that
\begin{align*}
X_{h} & \ge E_{h}OX_{g}O^{T}E_{h}\\
\iff E_{h}(X_{h}-OX_{g}O^{T})E_{h} & \ge0 & \because X_{h}=E_{h}X_{h}E_{h}\\
\iff E_{h}(X_{h}+c\mathbb{I}_{hg}-O(X_{g}-c\mathbb{I}_{hg})O^{T})E_{h} & \ge0\\
\iff X_{h}+c\mathbb{I}_{h} & \ge E_{h}O(X_{g}+c\mathbb{I}_{hg})O^{T}E_{h}, & \text{where }\mathbb{I}_{hg}:=\mathbb{I}.
\end{align*}
Further, 
\begin{align*}
X_{g}+c\mathbb{I}_{hg} & \ge X_{g}+c\mathbb{I}_{g}\\
\iff E_{h}O(X_{g}+c\mathbb{I}_{hg})O^{T}E_{h} & \ge E_{h}O(X_{g}+c\mathbb{I}_{g})O^{T}E_{h}
\end{align*}
which together yield 
\[
X_{h}\ge E_{h}OX_{g}O^{T}E_{h}\iff X_{h}+c\mathbb{I}_{h}\ge E_{h}O(X_{g}+c\mathbb{I}_{g})O^{T}E_{h}.
\]

\end{proof}

\begin{lem}
\label{lem:spanningLemma}Consider an $n$-dimensional vector space.
Given a diagonal matrix $X=\diag(x_{1},x_{2}\dots x_{n})$ and a vector
$\left|c\right\rangle =(c_{1},c_{2}\dots,c_{n})$ where all the $x_{i}$s
are distinct and all the $c_{i}$ are non-zero, the vectors $\left|c\right\rangle ,X\left|c\right\rangle ,\dots X^{n-1}\left|c\right\rangle $
span the vector space.
\end{lem}

\begin{proof}
We write the vectors as 
\[
\left|\tilde{w}_{i}\right\rangle =X^{i-1}\left|c\right\rangle =\left[\begin{array}{c}
x_{1}^{i-1}c_{1}\\
x_{2}^{i-1}c_{2}\\
\vdots\\
x_{n}^{i-1}c_{n}
\end{array}\right].
\]
We show that the set of vectors are linearly independent, which is
equivalent to showing that the determinant of the matrix containing
the vectors as rows (or equivalently as columns) is non-zero, i.e.
\[
\det\left(\underbrace{\left[\begin{array}{ccccc}
1 & 1 & \dots &  & 1\\
x_{1} & x_{2} &  &  & x_{n}\\
x_{1}^{2} & x_{2}^{2} &  &  & x_{n}^{2}\\
\vdots &  & \ddots\\
x_{1}^{n-1} & x_{2}^{n-1} & \dots &  & x_{n}^{n-1}
\end{array}\right]}_{:=\tilde{X}}\left[\begin{array}{ccccc}
c_{1}\\
 & c_{2}\\
\\
 &  &  & \ddots\\
 &  &  &  & c_{n}
\end{array}\right]\right)=c_{1}\cdot c_{2}\cdot\dots c_{n}\cdot\det\tilde{X}
\]
is non-zero. To see this, we note that $\tilde{X}$ is the so-called
Vandermonde matrix (restricted to being a square matrix) and its determinant,
known as the Vandermonde determinant, is $\det(\tilde{X})=\prod_{1\le i\le j\le n}(x_{j}-x_{i})\neq0$
as $x_{i}$s are distinct. As $c_{i}$s are all non-negative, this
concludes the proof.
\end{proof}

\begin{lem}
Let $t=\sum_{i=1}^{n}p_{i}\left[ x_{i}\right] $
be the zeroth assignment for a set of real numbers $0\le x_{1}<x_{2}\dots<x_{n}$.
Then for $0\le k\le n-2$, 
$$\left\langle x^{k}\right\rangle =0\ \ \ \
\text{and}\ \ \ 
\left\langle x^{n-1}\right\rangle >0,$$
where $\left\langle x^{k}\right\rangle =\sum_{i=1}^{n}p_{i}\left(x_{i}\right)^{k}$.
\label{lem:expectationLemma}
\end{lem}
\begin{proof}
For the proof, see Section 4 and Appendix B of \cite{Arora2018}. Most of the work had already been done by Mochon \cite{Mochon07}.
\end{proof}

\section{Proofs and examples for balanced monomial assignments}
\label{app:BalancedMon}

\subsection*{Proof of \Propref{ExactSolnBalancedMonomialAligned}}

{\begin{proof}
The orthonormal basis (over $\text{span}\{\left|h_{1}\right\rangle ,\left|h_{2}\right\rangle \dots\left|h_{n}\right\rangle \}$)
of interest here is
\begin{equation}
\left|w'_{i}\right\rangle :=\frac{\Pi_{h_{i}}^{\perp}(X_{h})^{i}\left|w'\right\rangle }{\sqrt{c_{h_{i}}}}\label{eq:alignedBalancedMonomialVectors}
\end{equation}
which entails 
\begin{equation}
\Pi_{h_{i}}^{\perp}=\begin{cases}
\mathbb{I}_{h} & i=0\\
\mathbb{I}_{h}-\sum_{j=i+1}^{0}\left|w'_{j}\right\rangle \left\langle w'_{j}\right| & i<0\\
\mathbb{I}_{h}-\sum_{j=-b}^{i-1}\left|w'_{j}\right\rangle \left\langle w'_{j}\right| & i>0
\end{cases}\label{eq:alignedBalancedMonomialProjectors}
\end{equation}
where $\mathbb{I}_{h}:=E_{h}$. We define $\left|v'_{i}\right\rangle $
and $\Pi_{g_{i}}^{\perp}$ analogously. Our strategy would be to keep
track of both the highest and lowest power $l$, in $\left\langle w'\right|X_{h}^{l}\left|w'\right\rangle $
and $\left\langle v'\right|X_{g}^{l}\left|v'\right\rangle $, which
appear in the matrix elements $\left\langle w'_{i}\right|D\left|w'_{j}\right\rangle $.
We use $\left\langle x_{h}^{l}\right\rangle ^{\prime}:=\left\langle w'\right|X_{h}^{l}\left|w'\right\rangle =\left\langle w\right|X_{h}^{l+2b}\left|w\right\rangle $
and similarly $\left\langle x_{g}^{l}\right\rangle ^{\prime}:=\left\langle v'\right|X_{g}^{l}\left|v'\right\rangle =\left\langle w\right|X_{g}^{l+2b}\left|w\right\rangle $.
To this end, we denote the minimum and maximum powers $l$, by 
\[
\mathcal{M}(\left|w'_{i}\right\rangle )=\begin{cases}
\minmax{\left\langle x_{h}^{0}\right\rangle ^{\prime}\left|w'\right\rangle }{\left\langle x_{h}^{0}\right\rangle ^{\prime}\left|w'\right\rangle } & i=0\\
\minmax{\left\langle x_{h}^{-2|i|}\right\rangle ^{\prime}(X_{h})^{-|i|}\left|w'\right\rangle }{\left\langle x_{h}^{0}\right\rangle ^{\prime}\left|w'\right\rangle } & i<0\\
\minmax{\left\langle x_{h}^{-2b}\right\rangle ^{\prime}(X_{h})^{-b}\left|w'\right\rangle }{\left\langle x_{h}^{2i}\right\rangle ^{\prime}(X_{h})^{i}\left|w'\right\rangle } & i>0.
\end{cases}
\]

We define $D:=X_{h}-E_{h}OX_{g}O^{T}E_{h}\doteq\left\langle w'_{i}\right|\left(X_{h}-E_{h}OX_{g}O^{T}E_{h}\right)\left|w_{j}'\right\rangle $.
It suffices to restrict to the span of $\{\left|w_{i}'\right\rangle \}$
basis, because $X_{h}\left|v'_{i}\right\rangle =0$ and $E_{h}\left|v'_{i}\right\rangle =0$.
The lowest power $l$, appearing in $D$ is for $i=j=-b$ (as $-b\le i,j\le n-b-1$).
This can be evaluated to be $-2b$ by observing that 
\[
\mathcal{M}(\left\langle w'_{-b}\right|)X_{h}\mathcal{M}(\left|w'_{-b}\right\rangle )=\minmax{\left\langle x_{h}^{-2b}\right\rangle ^{\prime}\left\langle x_{h}^{-2b}\right\rangle ^{\prime}\left\langle x_{h}^{-2b+1}\right\rangle ^{\prime}}{\left\langle x_{h}^0 \right\rangle ^{\prime}\left\langle x_{h}^0 \right\rangle ^{\prime}\left\langle x_{h}\right\rangle ^{\prime}},
\]
where we multiplied component-wise. To find the highest power $l$, in
the matrix $D$, note that for $i,j>0$ we have 
\[
\mathcal{M}(\langle w'_{i}|)X_{h}\mathcal{M}(|w'_{j}\rangle) =\minmax{\left\langle x_{h}^{-2b}\right\rangle ^{\prime}\left\langle x_{h}^{-2b}\right\rangle ^{\prime}\left\langle x_{h}^{-2b+1}\right\rangle ^{\prime}}{\left\langle x_{h}^{2i}\right\rangle ^{\prime}\left\langle x_{h}^{2j}\right\rangle ^{\prime}\left\langle x_{h}^{i+j+1}\right\rangle ^{\prime}},
\]
therefore $l=\max\{2i,2j,i+j+1\}$. As argued for the zeroth assignment
$l=2n-2b-1$ for $i=j=n-b-1$ or otherwise strictly less than $2n-2b-1$.
Thus, only the $D_{n-b-1,n-b-1}$ term in $D$ depends on $\left\langle x_{h}^{2n-2b-1}\right\rangle ^{\prime}$.
Except for this term, all other terms depend, at most, on $\left\langle x_{h}^{-2b}\right\rangle ^{\prime},\left\langle x_{h}^{-2b+1}\right\rangle ^{\prime},\dots\left\langle x_{h}^{2n-2b-2}\right\rangle ^{\prime}$,\\
i.e. $\left\langle x_{h}^{0}\right\rangle ,\left\langle x_{h}^{1}\right\rangle ,\dots\left\langle x_{h}^{2n-2}\right\rangle $.
The analogous argument for $\left\langle v'_{i}\right|X_{g}\left|v'_{j}\right\rangle $,
the observation that $\left\langle w'_{i}\right|D\left|w'_{j}\right\rangle =\left\langle w'_{i}\right|X_{h}\left|w'_{j}\right\rangle -\left\langle v'_{i}\right|X_{g}\left|v'_{j}\right\rangle $,
and the fact that $\left\langle x^{0}\right\rangle =\left\langle x^{1}\right\rangle =\dots=\left\langle x^{2n-2}\right\rangle =0$
entail that these terms vanish. It remains to establish that $D_{n-b-1,n-b-1}\ge0$.
This is easily seen by noting that in $\left\langle w'_{n-b-1}\right|D\left|w'_{n-b-1}\right\rangle $,
the only term which would not get cancelled due to the aforesaid reasoning,
must come from the part of $\left|w'_{n-b-1}\right\rangle $ containing
$X_{h}^{n-b-1}\left|w'\right\rangle $. It suffices to show that the
coefficient of this term is positive, as we know that $\left\langle x^{2n-2b-1}\right\rangle ^{\prime}=\left\langle x^{2n-1}\right\rangle >0$.
Further, from \Eqref{alignedBalancedMonomialProjectors} and \Eqref{alignedBalancedMonomialVectors}, we know that the coefficient is $1/c_{h_{n-b-1}}$.
This establishes $D\ge0$.
\end{proof}
\vspace{\baselineskip}
\subsection*{Example of balanced aligned and misaligned monomial assignments}

Let us consider a concrete example of a balanced
aligned monomial assignment with $2n=8$ and $m=2b=2$ (see \Figref{balancedAlignedmAssignment}).
We represent the range of dependence of $\left\langle w'_{0}\right|X_{h}\left|w'_{0}\right\rangle $
on $\left\langle x_{h}^{l}\right\rangle $ diagrammatically by enclosing
in a left bracket, the terms $\left\langle x^{3}\right\rangle =\left\langle x\right\rangle ^{\prime}$
and $\left\langle x^{2}\right\rangle =\left\langle x^{0}\right\rangle ^{\prime}$
(replacing $\left|w\right\rangle $ with $\left|w'_{0}\right\rangle $)
and writing $\left|w'_{0}\right\rangle $ next to it. Similarly, for
$\left|w'_{-1}\right\rangle ,\left|w'_{1}\right\rangle $ and $\left|w'_{2}\right\rangle $
we enclose in a left bracket, the terms 
\[
\left\{ \left\langle x^{0}\right\rangle ,\left\langle x^{1}\right\rangle ,\left\langle x^{2}\right\rangle ,\left\langle x^{3}\right\rangle \right\} =\left\{ \left\langle x^{-2}\right\rangle ^{\prime},\left\langle x^{-1}\right\rangle ^{\prime},\dots\left\langle x\right\rangle ^{\prime}\right\} ,
\]
\[
\left\{ \left\langle x^{0}\right\rangle ,\left\langle x^{1}\right\rangle ,\dots,\left\langle x^{5}\right\rangle \right\} =\left\{ \left\langle x^{-2}\right\rangle ^{\prime},\left\langle x^{-1}\right\rangle ^{\prime},\dots\left\langle x^{3}\right\rangle ^{\prime}\right\} 
\]
 and 
\[
\left\{ \left\langle x^{0}\right\rangle ,\left\langle x^{1}\right\rangle ,\dots\left\langle x^{7}\right\rangle \right\} =\left\{ \left\langle x^{-2}\right\rangle ^{\prime},\left\langle x^{-1}\right\rangle ^{\prime},\dots\left\langle x^{5}\right\rangle ^{\prime}\right\}, 
\]
 respectively. Note that the highest power $l$ of $\left\langle x_{h}^{l}\right\rangle $
that appears in $\left\langle w'_{i}\right|X_{h}\left|w'_{j}\right\rangle $
is $l=7$ only when $i=j=2$. Thus, the matrix $D$ restricted
to the subspace spanned by the $\{\left|w_{i}'\right\rangle \}$ basis
(again, we can safely ignore the subspace $\text{span}\{\left|v_{i}'\right\rangle \}$
because $D\left|v'_{i}\right\rangle =0$) has only one non-zero entry,
which is positive, as $\left\langle x^{7}\right\rangle >0$.

We now explain why a direct extension of the analysis to the balanced
misaligned monomial assignment fails and subsequently see how to remedy
the situation. Consider the case with $2n=8$ and
$m=2b-1=3$ (see \Figref{balancedMisalignedMassignment}). From hindsight,
we write both the $\left|v'_{i}\right\rangle $s and the $\left|w'_{i}\right\rangle $s.
We start with $\left|w'_{0}\right\rangle =X_{h}^{3/2}\left|w\right\rangle $
and $\left|v'_{0}\right\rangle =X_{g}^{3/2}\left|v_{0}\right\rangle $,
and, as before, enclose the terms $\left\{ \left\langle x^{0}\right\rangle ^{\prime}=\left\langle x^{3}\right\rangle ,\left\langle x^{1}\right\rangle ^{\prime}=\left\langle x^{4}\right\rangle \right\} $
in a left bracket. We continue by multiplying $\left|w'_{0}\right\rangle $
with $X_{h}^{-1}$ (and $\left|v'_{0}\right\rangle $ with $X_{g}^{-1}$,
respectively) and projecting out the components along the previous
vectors. We represent these by $\left|w_{-1}'\right\rangle $ and
$\left|v'_{-1}\right\rangle $ and in the figure, enclose the terms
$\left\{ \left\langle x\right\rangle =\left\langle x^{-2}\right\rangle ^{\prime},\left\langle x^{2}\right\rangle =\left\langle x^{-1}\right\rangle ^{\prime}\dots\left\langle x^{4}\right\rangle =\left\langle x\right\rangle ^{\prime}\right\} $
in the left and right brackets. We do not continue further, because in this case  a dependence on $\left\langle x^{-1}\right\rangle $ arises and persists for subsequent vectors. In general, we stop after taking
$b$ (which equals $1$ here) steps downwards. We can move upwards by multiplying
 $\left|w_{0}'\right\rangle $ with $X_{h}$ (and $\left|v_{0}'\right\rangle $
with $X_{g}$ resp.) and projecting out the components along the previous
vectors. We represent these by $\left|w_{1}'\right\rangle $ and $\left|v_{1}'\right\rangle $ and in
the figure, enclose the terms $\left\{ \left\langle x\right\rangle =\left\langle x^{-2}\right\rangle ^{\prime},\left\langle x^{2}\right\rangle =\left\langle x^{-1}\right\rangle ^{\prime}\dots\left\langle x^{6}\right\rangle =\left\langle x^{3}\right\rangle ^{\prime}\right\} $
in the brackets. Finally, we construct $\left|w'_{2}\right\rangle $
and $\left|v'_{2}\right\rangle $ by taking a step up using $X_{h}$
and $X_{g}$, respectively (these are essentially fixed to be the vectors
orthogonal to the previous ones once we restrict to $\text{span}(\left|h_{1}\right\rangle ,\left|h_{2}\right\rangle \dots\left|h_{n}\right\rangle )$)
and $\text{span}(\left|g_{1}\right\rangle ,\left|g_{2}\right\rangle \dots\left|g_{n}\right\rangle )$).
Taking a step down using $X_{h}^{-1}$ and $X_{g}^{-1}$ we could
have constructed $\left|w'_{-2}\right\rangle $ and $\left|v'_{-2}\right\rangle $
respectively but they are the same as $\left|w'_{2}\right\rangle $
and $\left|v'_{2}\right\rangle $. If
we were to use $O=\sum_{i=-1}^{2}\left(\left|w'_{i}\right\rangle \left\langle v'_{i}\right|+\hc\right)$, we would have obtained dependence on $\left\langle x^{7}\right\rangle $
in the last row (corresponding to $\left|w'_{2}\right\rangle $) and
a dependence on $\left\langle x^{8}\right\rangle $ for the last term
(i.e. $\left\langle w'_{2}\right|D\left|w'_{2}\right\rangle $). This
already hints that the matrix is negative because it has the form
$\left[\begin{array}{cc}
0 & b\\
b & c
\end{array}\right]$ with $b\neq0$, which means that the determinant is $-b^{2}$, entailing
there's a negative eigenvalue; thus this choice can not work. We therefore
define $O:=\left(\sum_{i=-1}^{1}\left|w'_{i}\right\rangle \left\langle v'_{i}\right|+\hc\right)+\left|w'_{2}\right\rangle \left\langle w'_{2}\right|+\left|v'_{2}\right\rangle \left\langle v'_{2}\right|$. Furthermore, instead of using 
\begin{equation}
X_{h}\ge E_{h}OX_{g}O^{T}E_{h}\label{eq:balancedMisalignedCaseEx}
\end{equation}
 for establishing positivity, we equivalently use 
\begin{equation}
E_{h}\ge\left(X_{h}^{\dashv}\right)^{1/2}OX_{g}O^{T}\left(X_{h}^{\dashv}\right)^{1/2}.\label{eq:balancedMisalignedInvertedEx}
\end{equation}
The reason is that to establish positivity, we must include $\left|w'_{2}\right\rangle $
in the basis (we can neglect the null vectors of $E_{h}$), and even
though the RHS of \Eqref{balancedMisalignedCaseEx} would not contribute,
the LHS would get non-trivial contributions along the rows (as was
the case earlier). Using the form with the inverses lets us remove
this dependence. To see this, note that $\text{span}\{\left|w'_{-1}\right\rangle ,\left|w'_{0}\right\rangle \dots\left|w'_{2}\right\rangle \}$
equals the $h$-space, i.e. $\text{span}\{\left|h_{1}\right\rangle ,\left|h_{2}\right\rangle \dots\left|h_{n}\right\rangle \}$.
Further, $\text{span}\{X_{h}^{1/2}\left|w'_{i}\right\rangle \}_{i=-1}^{2}$
also equals the $h$-space (but the vectors are not, in general, orthonormal
any more). Finally, observe that $X_{h}^{1/2}\left|w'_{2}\right\rangle $
is a null vector of the RHS of \Eqref{balancedMisalignedInvertedEx}.
Therefore, to prove the positivity, it suffices to restrict
to $\text{span}\{X_{h}^{1/2}\left|w'_{i}\right\rangle \}_{i=-1}^{1}$.
An arbitrary normalised vector in this space can be written as 
\begin{align*}
\left|\psi\right\rangle  & =\frac{\sum_{i=-1}^{1}\alpha_{i}X_{h}^{1/2}\left|w_{i}'\right\rangle }{\sqrt{\sum_{i,j=-1}^{1}\alpha_{i}\alpha_{j}\left\langle w'_{i}\right|X_{h}\left|w'_{j}\right\rangle }}\\
\implies X_{g}^{1/2}O^{T}(X_{h}^{\dashv})^{1/2}\left|\psi\right\rangle  & =\frac{\sum_{i=-1}^{1}\alpha_{i}X_{g}^{1/2}\left|v'_{i}\right\rangle }{\sqrt{\sum_{i,j=-1}^{1}\alpha_{i}\alpha_{j}\left\langle w'_{i}\right|X_{h}\left|w'_{j}\right\rangle }}\\
\implies\left\langle \psi\right|(X_{h}^{\dashv})^{1/2}OX_{g}O^{T}(X_{h}^{\dashv})^{1/2}\left|\psi\right\rangle  & =\frac{\sum_{i,j=-1}^{1}\alpha_{i}\alpha_{j}\left\langle v_{i}'\right|X_{g}\left|v_{j}'\right\rangle }{\sum_{i,j=-1}^{1}\alpha_{i}\alpha_{j}\left\langle w'_{i}\right|X_{h}\left|w'_{j}\right\rangle }=1,
\end{align*}
where we got the equality by noting that $\left\langle v'_{i}\right|X_{g}\left|v'_{j}\right\rangle $s
depend on (at most) $\left\{ \left\langle x_{g}\right\rangle ,\left\langle x_{g}^{2}\right\rangle \dots\left\langle x_{g}^{6}\right\rangle \right\} $
and analogously $\left\langle w'_{i}\right|X_{h}\left|w'_{j}\right\rangle $
depend on (at most) $\left\{ \left\langle x_{h}\right\rangle ,\left\langle x_{h}^{2}\right\rangle \dots\left\langle x_{h}^{6}\right\rangle \right\} $,
concluding that they are the same as $\left\langle x^{i}\right\rangle =0$
for $i\in\{0,1,\dots6\}$. Since we proved that the RHS of \Eqref{balancedMisalignedInvertedEx}
is one for all normalised $\left|\psi\right\rangle $s, we infer that
we have the correct orthogonal matrix. 

\begin{figure}
\noindent \begin{centering}
\hfill{}\subfloat[$2n=8$, $m=2b=2$; Balanced (aligned) $m$-assignment. \label{fig:balancedAlignedmAssignment}]{\noindent \begin{centering}
\includegraphics{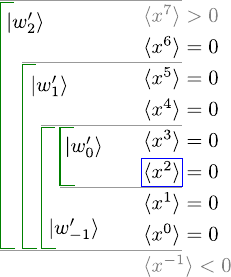}
\par\end{centering}
}\hfill{}\subfloat[$2n=8$, $m=2b-1=3$; Balanced (aligned) monomial assignment. \label{fig:balancedMisalignedMassignment}]{\noindent \begin{centering}
\includegraphics{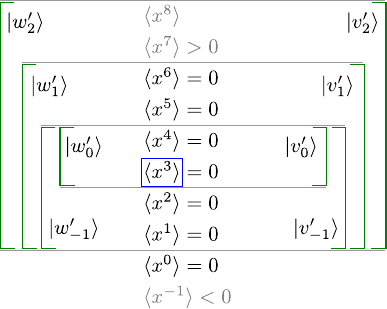}
\par\end{centering}
}\hfill{}
\par\end{centering}
\caption{Depicting balanced monomial assignments with simple examples.}
\end{figure}

\vspace{\baselineskip}

\subsection*{Proof of \Propref{ExactSolnBalancedMonomialMisaligned}}
\begin{proof}
The proof is very similar to that of \Propref{ExactSolnBalancedMonomialAligned}.
 The orthonormal basis (over $\left\{ \left|h_{1}\right\rangle ,\left|h_{2}\right\rangle \dots\left|h_{n}\right\rangle \right\} $)
of interest here is 
\[
\left|w'_{i}\right\rangle :=\frac{\Pi_{h_{i}}^{\perp}(X_{h})^{i}\left|w'\right\rangle }{\sqrt{c_{h_{i}}}}
\]
which entails 
\[
\Pi_{h_{i}}^{\perp}=\begin{cases}
\mathbb{I}_{h} & i=0\\
\mathbb{I}_{h}-\sum_{j=i-1}^{0}\left|w'_{j}\right\rangle \left\langle w'_{j}\right| & i<0\\
\mathbb{I}_{h}-\sum_{j=-b+1}^{i}\left|w'_{j}\right\rangle \left\langle w'_{j}\right| & i>0
\end{cases},
\]
where $\mathbb{I}_{h}:=E_{h}$. We define $\left|v'_{i}\right\rangle $
and $\Pi_{g_{i}}^{\perp}$ analogously. Our strategy is to keep track
of the highest and lowest powers $l$ in $\left\langle w'\right|X_{h}^{l}\left|w'\right\rangle $
and $\left\langle v'\right|X_{g}^{l}\left|v'\right\rangle $, which
appear in the matrix elements $\left\langle w_{i}'\right|X_{h}\left|w'_{j}\right\rangle $
and $\left\langle v'_{i}\right|X_{g}\left|v'_{j}\right\rangle $.
For brevity, as before, we use $\left\langle x_{h}^{l}\right\rangle ^{\prime}:=\left\langle w'\right|X_{h}^{l}\left|w'\right\rangle $
and similarly $\left\langle x_{g}^{l}\right\rangle ^{\prime}:=\left\langle v'\right|X_{g}^{l}\left|v'\right\rangle $.
To this end, we denote the minimum and maximum powers $l$, by 
\[
\mathcal{M}(\left|w'_{i}\right\rangle )=\begin{cases}
\left(\left\langle x_{h}^{0}\right\rangle ^{\prime}\left|w'\right\rangle ,\left\langle x_{h}^{0}\right\rangle ^{\prime}\left|w'\right\rangle \right) & i=0\\
\left(\left\langle x_{h}^{-2\left|i\right|}\right\rangle ^{\prime}(X_{h})^{-\left|i\right|}\left|w'\right\rangle ,\left\langle x_{h}^{0}\right\rangle ^{\prime}\left|w'\right\rangle \right) & i<0\\
\left(\left\langle x_{h}^{-2(b-1)}\right\rangle ^{\prime}(X_{h})^{-(b-1)}\left|w'\right\rangle ,\left\langle x_{h}^{2i}\right\rangle ^{\prime}(X_{h})^{i}\left|w'\right\rangle \right) & i>0.
\end{cases}
\]
Note that establishing $X_{h}\ge E_{h}OX_{g}O^{T}E_{h}$ is equivalent
to establishing 
\begin{equation}
E_{h}\ge X_{h}^{-1/2}OX_{g}O^{T}X_{h}^{-1/2}.\label{eq:invertedBalancedMisaligned}
\end{equation}

It is easy to see that $X_{h}^{1/2}\left|w'_{n-b}\right\rangle $
is a null vector (vector with zero eigenvalue) for the RHS as $X_{g}O^T \left|w'_{n-b}\right\rangle =0$.
Any vector $\left|\psi\right\rangle $ in $\text{span}\{\left|g_{1}\right\rangle ,\left|g_{2}\right\rangle \dots\left|g_{n}\right\rangle \}$
is a null vector for both the LHS and the RHS. Thus, we can restrict to $\text{span}\{\left|h_{1}\right\rangle ,\left|h_{2}\right\rangle ,\dots\left|h_{n}\right\rangle \}\backslash\text{span}\{X_{h}^{1/2}\left|w'_{n-b}\right\rangle \}$,
i.e. to vectors in the $h$-space orthogonal to $X_{h}^{1/2}\left|w'_{n-b}\right\rangle $, in order to establish positivity.
It turns out to be easier to test for positivity on a possibly larger
space. It is clear that $\text{span}\left\{ X_{h}^{1/2}\left|w'_{i}\right\rangle \right\} _{i=-b+1}^{n-b}=\text{span}\{\left|h_{1}\right\rangle ,\left|h_{2}\right\rangle \dots\left|h_{n}\right\rangle \}$
(because it also equals $\text{span}\{\left|w'\right\rangle _{i}\}_{i=-b+1}^{n-b}$,
due to \Lemref{spanningLemma}). As neglecting vectors with components
along $X_{h}^{1/2}\left|w'_{n-b}\right\rangle $ suffices for establishing
positivity of \Eqref{invertedBalancedMisaligned}, we can restrict
to $\text{span}\{X_{h}^{1/2}\left|w'_{i}\right\rangle \}_{i=-b+1}^{n-b-1}$, which might still contain vectors with components along $X_{h}^{1/2}\left|w'_{n-b}\right\rangle $,
as the basis vectors are not orthogonal. 
Let $\left|\psi\right\rangle =\left(\sum_{i=-b+1}^{n-b-1}\alpha_{i}X_{h}^{1/2}\left|w'_{i}\right\rangle \right)/c$
where $c=\sqrt{\left\langle \psi|\psi\right\rangle }$. To establish
\Eqref{invertedBalancedMisaligned}, it is enough to show that for
all choices of $\alpha_{i}$s, 
\begin{align}
1 & \ge\left\langle \psi\right|X_{h}^{-1/2}OX_{g}O^{T}X_{h}^{-1/2}\left|\psi\right\rangle \nonumber \\
 & =\frac{\sum_{i,j=-b+1}^{n-b-1}\alpha_{i}\alpha_{j}\left\langle v'_{i}\right|X_{g}\left|v'_{j}\right\rangle }{\sum_{i,j=-b+1}^{n-b-1}\alpha_{i}\alpha_{j}\left\langle w'_{i}\right|X_{h}\left|w'_{j}\right\rangle }\label{eq:ratioForInequality}\\
 & =1\nonumber, 
\end{align}
where the second step follows from the fact that $X_{g}^{1/2}O^{T}X_{h}^{-1/2}\left|\psi\right\rangle =\sum_{i=-b+1}^{n-b-1}\alpha_{i}X_{g}^{1/2}\left|v'_{i}\right\rangle $,
and the last step follows from a counting argument which we give below.

Note that 
\begin{equation}
\left\langle x_{h}^{i}\right\rangle ^{\prime}=\left\langle x_{h}^{i+2b-1}\right\rangle\nonumber \label{eq:monomialmisalignedMochonPowerShifting}
\end{equation}
 and 
 
\begin{equation}
\left\langle x^{0}\right\rangle =\left\langle x\right\rangle =\dots=\left\langle x^{2n-2}\right\rangle =0.\label{eq:monomialmisalignedMochonPowers}
\end{equation}
To determine the highest power $l$ in $\left\langle w'\right|X_{h}^{l}\left|w'\right\rangle $
which appears in the matrix elements $\left\langle w'_{i}\right|X_{h}\left|w'_{j}\right\rangle $
(for $-b+1\le i,j\le n-b-1$) it suffices to consider $\left\langle w'_{n-b-1}\right|X_{h}\left|w'_{n-b-1}\right\rangle $.
To this end, we evaluate 
\begin{align*}
 & \mathcal{M}(\left\langle w'_{n-b-1}\right|)X_{h}\mathcal{M}(\left|w'_{n-b-1}\right\rangle )\\
 & =\left(\left\langle x_{h}^{-2(b-1)}\right\rangle ^{\prime}\left\langle x_{h}^{-2(b-1)}\right\rangle ^{\prime}\left\langle x_{h}^{-2(b-1)+1}\right\rangle ^{\prime},\left\langle x_{h}^{2(n-b-1)}\right\rangle ^{\prime}\left\langle x_{h}^{2(n-b-1)}\right\rangle ^{\prime}\left\langle x_{h}^{2(n-b-1)+1}\right\rangle ^{\prime}\right)\\
 & =\left(\left\langle x_{h}\right\rangle \left\langle x_{h}\right\rangle \left\langle x_{h}^{2}\right\rangle ,\left\langle x_{h}^{2n-3}\right\rangle \left\langle x_{h}^{2n-3}\right\rangle \left\langle x_{h}^{2n-2}\right\rangle \right).
\end{align*}
The highest power is $l=2n-2$. To find the lowest power
of $l$ in $\left\langle w'\right|X_{h}^{l}\left|w'\right\rangle $
which appears in the matrix elements $\left\langle w'_{i}\right|X_{h}\left|w'_{j}\right\rangle $
(for $-b+1\le i,j\le n-b-1$) it suffices to consider $\left\langle w'_{-b+1}\right|X_{h}\left|w'_{-b+1}\right\rangle $.
To this end, we evaluate 
\begin{align*}
\mathcal{M}(\left\langle w'_{-b+1}\right|)X_{h}\mathcal{M}(\left|w'_{-b+1}\right\rangle ) & =\left(\left\langle x_{h}^{-2(b-1)}\right\rangle ^{\prime}\left\langle x_{h}^{-2(b-1)}\right\rangle ^{\prime}\left\langle x_{h}^{-2(b-1)+1}\right\rangle ^{\prime},\left\langle x_{h}^{0}\right\rangle ^{\prime}\left\langle x_{h}^{0}\right\rangle ^{\prime}\left\langle x_{h}\right\rangle ^{\prime}\right)\\
 & =\left(\left\langle x_{h}\right\rangle \left\langle x_{h}\right\rangle \left\langle x_{h}^{2}\right\rangle ,\left\langle x_{h}^{2b-1}\right\rangle \left\langle x_{h}^{2b-1}\right\rangle \left\langle x_{h}^{2b}\right\rangle \right).
\end{align*}
The lowest power is $l=1$. We thus conclude that the
numerator in \Eqref{ratioForInequality} is a function of $\left\langle x_{h}\right\rangle ,\left\langle x_{h}^{2}\right\rangle ,\dots\text{\ensuremath{\left\langle x_{h}^{2n-2}\right\rangle }}$, and analogously the denominator is a function of
$\left\langle x_{g}\right\rangle ,\left\langle x_{g}^{2}\right\rangle ,\dots\left\langle x_{g}^{2n-2}\right\rangle $
with the same form. Using \Eqref{monomialmisalignedMochonPowers},
we obtain that the numerator and the denominator are the same.
\end{proof}

\section{Proofs and examples for unbalanced monomial assignments}
\label{app:UnbalancedMon}
\subsection*{Proof of \Propref{ExactSolnUnbalancedMonomialAligned}}
\begin{proof}
Many observations from the proof of \Propref{ExactSolnBalancedMonomialAligned}
carry over to this case. We import the definitions of $\left\{ \left|w'_{i}\right\rangle \right\} _{i=-b}^{n-b-2}$
and $\{\left|v'_{i}\right\rangle \}_{i=-b}^{n-b-1}$, together with the
observations that $\mathcal{M}(\left\langle w'_{-b}\right|)X_{h}\mathcal{M}(\left|w'_{-b}\right\rangle )$
has no dependence on $\left\langle x_{h}^{l}\right\rangle ^{\prime}$
with $l$ smaller than $-2b$ (which corresponds to $\left\langle x_{h}\right\rangle $),
and that $\mathcal{M}(\left\langle w'_{n-b-2}\right|)X_{h}\mathcal{M}(\left|w'_{n-b-2}\right\rangle )$
has no dependence on $\left\langle x_{h}^{l}\right\rangle ^{\prime}$
with $l$ greater than $2n-2b-4+1=2n-3-2b$. We can restrict to $\text{span}\{\left|w'_{-b}\right\rangle ,\left|w'_{-b+1}\right\rangle \dots\left|w'_{n-b-2}\right\rangle \}$
to establish the positivity of $D:=X_{h}-E_{h}OX_{g}O^{T}E_{h}$.
Using the analogous observation for $\mathcal{M}(\left\langle v'_{-b}\right|)X_{g}\mathcal{M}(\left|v'_{-b}\right\rangle )$
and $\mathcal{M}(\left\langle v'_{n-b-2}\right|)X_{g}\mathcal{M}(\left|v'_{n-b-2}\right\rangle )$,
along with the fact that $\left\langle x^{l}\right\rangle ^{\prime}=\left\langle x^{l+2b}\right\rangle $
and $\left\langle x^{0}\right\rangle =\left\langle x^{1}\right\rangle =\dots=\left\langle x^{2n-3}\right\rangle =0$,
it follows that $D$ is zero.
\end{proof}

\vspace{\baselineskip}
\subsection*{Proof of \Propref{ExactSolnUnbalancedMonomialMisaligned}}

\begin{proof}
For this proof, we can use the definitions and observations from the
proof of \Propref{ExactSolnBalancedMonomialMisaligned}. We import
the definitions of $\left\{ \left|w'_{i}\right\rangle \right\} {}_{i=-b+1}^{n-b}$
and $\left\{ \left|v'_{i}\right\rangle \right\} _{i=-b+1}^{n-b-1}$
along with the observation that 
\[
\mathcal{M}(\left\langle w'_{-b+1}\right|)X_{h}\mathcal{M}(\left|w'_{-b+1}\right\rangle )
\]
 has no dependence on $\left\langle x_{h}^{l}\right\rangle ^{\prime}$
with $l$ smaller than $-2b+2$ (which corresponds to $\left\langle x_{h}\right\rangle $),
and 
\[
\mathcal{M}(\left\langle w'_{n-b-1}\right|)X_{h}\mathcal{M}(\left|w'_{n-b-1}\right\rangle )
\]
 has no dependence on $\left\langle x^{l}\right\rangle $ with
$l$ greater than $2n-2b-1$ (which corresponds to $\left\langle x_{h}^{2n-2}\right\rangle $, as $2n-2b-1+(2b-1)=2n-2$). From the previous
proof, we also have that establishing $X_{h}\ge E_{h}OX_{g}O^{T}E_{h}$ is equivalent
to establishing that 
\[
1\ge\frac{\sum_{i,j=-b+1}^{n-b-1}\alpha_{i}\alpha_{j}\left\langle v'_{i}\right|X_{g}\left|v'_{j}\right\rangle }{\sum_{i,j=-b+1}^{n-b-1}\alpha_{i}\alpha_{j}\left\langle w'_{i}\right|X_{h}\left|w'_{j}\right\rangle },
\]
for all real $\{\alpha_{i}\}_{i=-b+1}^{n-b-1}$. We know that $\left\langle x\right\rangle =\left\langle x^{2}\right\rangle =\dots=\left\langle x^{2n-3}\right\rangle =0$.
As we have dependence on $\left\langle x_{h}^{2n-2}\right\rangle $,
we can't conclude that the fraction is one. However, as we saw in
the proof of \Propref{ExactSolnBalancedMonomialAligned}, dependence
on $\left\langle x_{h}^{2n-2}\right\rangle $ in the denominator
only appears in the $\left\langle w'_{n-b-1}\right|X_{h}\left|w'_{n-b-1}\right\rangle $
term with the positive coefficient $1/c_{h_{n-b-1}}$.
The analogous statement holds for the numerator. This, using $\left\langle x^{2n-2}\right\rangle >0$,
entails that the denominator is larger than or equal to the numerator,
concluding the proof.
\end{proof}
\vspace{\baselineskip}

\subsection*{Examples of unbalanced aligned and misaligned monomial assignments}

We illustrate how
the solution is constructed by considering a concrete example of an
unbalanced aligned monomial assignment. We start with $2n-1=7$
points and $m=2b=2$ (see \Figref{EvenUnbalancedMassignment}). We
use the same diagrammatic representation as before. In this case,
we have $4$ initial and $3$ final points and the basis
is $\left\{ \left|g_{1}\right\rangle ,\left|g_{2}\right\rangle ,\dots\left|g_{4}\right\rangle ,\left|h_{1}\right\rangle ,\left|h_{2}\right\rangle ,\left|h_{3}\right\rangle \right\} $.
We construct the basis of interest by starting at $\left|w'\right\rangle $
 and using $X_{h}^{-1}$
first until we reach $\left\langle x^{0}\right\rangle $, followed
by using $X_{h}$ until the space is spanned (analogously for $\left|v'\right\rangle $). We get $\left\{ \left|v'_{-1}\right\rangle ,\left|v'_{0}\right\rangle ,\left|v'_{1}\right\rangle ,\left|v'_{2}\right\rangle \right\} $
and $\left\{ \left|w'_{-1}\right\rangle ,\left|w'_{0}\right\rangle ,\left|w'_{1}\right\rangle \right\} $.
In the same vein as the previous solutions, we define $O:=\sum_{i=-1}^{1}\left(\left|w'_{i}\right\rangle \left\langle v'_{i}\right|+\hc\right)+\left|v'_{2}\right\rangle \left\langle v'_{2}\right|$.
In $X_{h}\ge E_{h}OX_{g}O^{T}E_{h}$, the $\left|v'_{2}\right\rangle $
term is removed by the projector $E_{h}:=\sum_{i=1}^{3}\left|h_{i}\right\rangle \left\langle h_{i}\right|$.
Using $\left\langle x^{0}\right\rangle =\left\langle x\right\rangle =\dots=\left\langle x^{5}\right\rangle =0$
and the counting arguments from before, it follows that $D=X_{h}-E_{h}OX_{g}O^{T}E_{h}$
is zero.

We now move on to
unbalanced misaligned monomial assignment. Consider $2n-1=7$ points
and $m=2b-1=1$. In this case, we have $3$ initial and $4$
final points and the basis is $\left\{ \left|g_{1}\right\rangle ,\left|g_{2}\right\rangle ,\left|g_{3}\right\rangle ,\left|h_{1}\right\rangle ,\left|h_{2}\right\rangle ,\dots\left|h_{4}\right\rangle \right\} $.
We construct the basis of interest by starting at
$\left|w'\right\rangle $ 
and using $X_{h}$ until the space is spanned (analogously for $\left|v'\right\rangle$). That is, we
first go downwards for $b-2$ steps (which is zero in this case), until
$\left\langle x\right\rangle $ is reached in the diagram. The basis
is $\left\{ \left|v'_{0}\right\rangle ,\left|v'_{1}\right\rangle ,\left|v'_{2}\right\rangle \right\} $
and $\left\{ \left|w'_{0}\right\rangle ,\left|w'_{1}\right\rangle ,\left|w'_{2}\right\rangle ,\left|w'_{3}\right\rangle \right\} $.
As before, we define $O:=\sum_{i=0}^{2}\left(\left|w'_{i}\right\rangle \left\langle v'_{i}\right|+\hc\right)+\left|w'_{3}\right\rangle \left\langle w'_{3}\right|$.
This time we
use $E_{h}\ge X_{h}^{-1/2}OX_{g}O^{T}X_{h}^{-1/2}$ which is equivalent
to $X_{h}\ge E_{h}OX_{g}O^{T}E_{h}$ for $E_{h}:=\sum_{i=1}^{4}\left|h_{i}\right\rangle \left\langle h_{i}\right|$.
Using an argument similar to the balanced misaligned case, we can reduce the
positivity condition to 
\[
1\ge\frac{\sum_{i,j=0}^{2}\alpha_{i}\alpha_{j}\left\langle v'_{i}\right|X_{g}\left|v'_{j}\right\rangle }{\sum_{i,j=0}^{2}\alpha_{i}\alpha_{j}\left\langle w'_{i}\right|X_{h}\left|w'_{j}\right\rangle },
\]
 but the counting argument doesn't make the fraction $1$. This is
because we now have an $\left\langle x_{h}^{6}\right\rangle $ dependence
in the denominator and $\left\langle x_{g}^{6}\right\rangle $ dependence
in the numerator. However, we also know that this term only appears
in $\left\langle w'_{2}\right|X_{h}\left|w'_{2}\right\rangle $ that
too with a positive coefficient. Furthermore, we know $\left\langle x_{h}^{6}\right\rangle >\left\langle x_{g}^{6}\right\rangle $
and therefore we can conclude that the numerator is smaller than the
denominator ensuring the inequality is always satisfied.

\begin{figure}[h!]
\hfill{}\subfloat[$2n-1=7;$ $m=2b=2$. Even unbalanced monomial assignment.\label{fig:EvenUnbalancedMassignment}]{\noindent \begin{centering}
\includegraphics{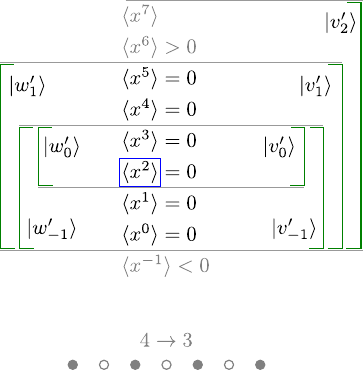}
\par\end{centering}
}\hfill{}\subfloat[$2n-1=7$; $m=2b-1=1$. Odd unbalanced monomial assignment.\label{fig:OddUnbalancedMassignment}]{\noindent \begin{centering}
\includegraphics{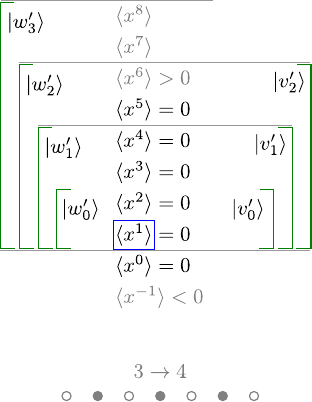}
\par\end{centering}
}\hfill{}
\caption{Depicting unbalanced monomial assignment with simple examples.}
\end{figure}

\section{Constructing a WCF protocol approaching bias 1/14}\label{app:example}
In this last part of the appendix we show how one can construct an explicit WCF protocol, in particular a protocol approaching bias $\epsilon=\frac{1}{14}$, corresponding to the point game with the same bias, that is for $k=3$ in $\epsilon(k)=\frac{1}{4k+2}$, we obtain $\epsilon(3)=\frac{1}{14}$. Several results and techniques presented in previous works, such as \cite{Mochon05,Mochon07, Aharon2014,Arora2019}, are required for this construction. We will only refer to them when they are needed.

The TDPG with bias $\frac{1}{14}$ includes the basic moves we mentioned in  Section \ref{sec:Preliminaries}, namely the split, merge and raise moves, as well as the main moves which are needed for the so-called \emph{ladder}, as  illustrated in Figure 3. %
We only need to determine the orthogonal matrix $O$ for these main moves, as the matrices corresponding to the split and the merge moves are given by the so-called \emph{blinkered unitary}, as presented in Equation 3 of \cite{Arora2019}, and the raise move is trivial, as it just increases the coordinate. 
The weights on the points constituting the ladder are given by the $f$-assignment. For our example (the bias $\frac{1}{14}$ case), the $f$-assignment is on a set of points seven points $\{x_{0}',x_{1}'\dots x_{6}'\}$, and the corresponding polynomial has degree five which we write as $f'(x)=(r_{1}'-x)(r_{2}'-x)(r_{3}'-x)(r_{4}'-x)(r_{5}'-x)$. More explicitly, the $f$-assignment is given by
$$
t'=\sum_{i=0}^{6}\frac{-f'(x_{i}')}{\prod_{j\neq i}(x_{j}'-x_{i}')}\left\llbracket x_{i}'\right\rrbracket .
$$
The placement of the roots of the polynomial with respect to the points is the following (see also Figure 3):
$$
x_{0}'=0<r_{1}'<r_{2}'<x_{1}'<x_{2}'<x_{3}'<x_{4}'<x_{5}'<x_{6}'<r_{3}'<r_{4}'<r_{5}'.$$

The assignment $t'$ includes a point with zero coordinate, while the orthogonal matrices $O$ (in \Propref{ExactSolnBalancedMonomialAligned},
\Propref{ExactSolnBalancedMonomialMisaligned}, \Propref{ExactSolnUnbalancedMonomialAligned},
and \Propref{ExactSolnUnbalancedMonomialMisaligned}) solve (monomial) assignments whose points have strictly positive coordinates. As already mentioned in \Secref{Mochon's-Assignments}, this is not really a restriction, as \Lemref{OriginIssueHandled} permits us to alternatively consider an $f$-assignment on the points $\{x_{0},x_{1}\dots x_{6}\}$
where $x_{i}=x_{i}'+c$ and $f(x)=(r_{1}-x)(r_{2}-x)\dots(r_{5}-x)$
where $r_{i}=r_{i}'+c$, for a positive number $c$. The resulting assignment 
$$
t=\sum_{i=0}^{6}\frac{-f(x_{i})}{\prod_{j\neq i}(x_{j}-x_{i})}\left\llbracket x_{i}\right\rrbracket$$
has the same solution as that of $t'$. We decompose $t$
into a sum of monomial assignments as
\begin{align*}
t & =\underbrace{\sum_{i=0}^{6}\frac{-r_{1}r_{2}r_{3}r_{4}r_{5}}{\prod_{j\neq i}(x_{j}-x_{i})}\left\llbracket x_{i}\right\rrbracket }_{\text{I}}+\underbrace{\sum_{i=0}^{6}\frac{-\overbrace{(r_{2}r_{3}r_{4}r_{5}+r_{1}r_{3}r_{4}r_{5}+r_{1}r_{2}r_{3}r_{5}+r_{1}r_{2}r_{3}r_{4})}^{:=\alpha_{1}}(-x_{i})}{\prod_{j\neq i}(x_{j}-x_{i})}\left\llbracket x_{i}\right\rrbracket }_{\text{II}}\\
 & \quad+\underbrace{\sum_{i=0}^{6}\frac{-\alpha_{2}(-x_{i})^{2}}{\prod_{j\neq i}(x_{j}-x_{i})}\left\llbracket x_{i}\right\rrbracket }_{\text{III}}+\underbrace{\sum_{i=0}^{6}\frac{-\alpha_{3}(-x_{i})^{3}}{\prod_{j\neq i}(x_{j}-x_{i})}\left\llbracket x_{i}\right\rrbracket }_{\text{IV}}\\
 & \quad+\underbrace{\sum_{i=0}^{6}\frac{-\alpha_{4}(-x_{i})^{4}}{\prod_{j\neq i}(x_{j}-x_{i})}\left\llbracket x_{i}\right\rrbracket }_{\text{V}}+\underbrace{\sum_{i=0}^{6}\frac{-\alpha_{5}(-x_{i})^{5}}{\prod_{j\neq i}(x_{j}-x_{i})}\left\llbracket x_{i}\right\rrbracket }_{\text{VI}},
\end{align*}
where $\alpha_{l}$ is the coefficient of $(-x)^{l}$ in $f(x)$.
Since the total number of points in each term is $7$, the monomial assignments are unbalanced. Terms I, III and V each have
an even powered monomial, therefore they correspond to the aligned
case. Their solutions, thus, are readily obtained from \Propref{ExactSolnUnbalancedMonomialAligned}.
Analogously, the remaining terms II, IV and VI have an odd powered
monomial, therefore they correspond to the misaligned case. Their solutions,
thus, are readily obtained from \Propref{ExactSolnUnbalancedMonomialMisaligned}.

We have already done  the hard work, which is to find the matrices which (effectively) solve the $f-$assignments for each move of the point game, and
we can now describe how the pieces fit together
to give the WCF protocol. We outline the steps of the associated TDPG, since, using the TEF, they can be seen as a short-hand to denote
an exchange and manipulation of quantum systems (e.g. qubits) by the two parties executing the WCF protocol, granted that
the associated unitaries are known (for details, see the description of the TEF in \cite{Arora2019}). Then, the WCF protocol consists of the same steps implemented in the reverse order. 
Here, we should clarify that, in fact, we convert a TIPG approaching bias $\frac{1}{14}$, into a TDPG following the technique presented, for instance, in the proof of  Theorem 5 in \cite{Aharon2014} with the minor modifications we outlined in Appendix \ref{app:sum-of-valid-functions}. Being familiar with the relationship between TIPGs and TDPGs and the related techniques facilitates the understanding of the construction that follows.

\begin{figure}[h]
  \begin{centering}
  \includegraphics[scale=0.8]{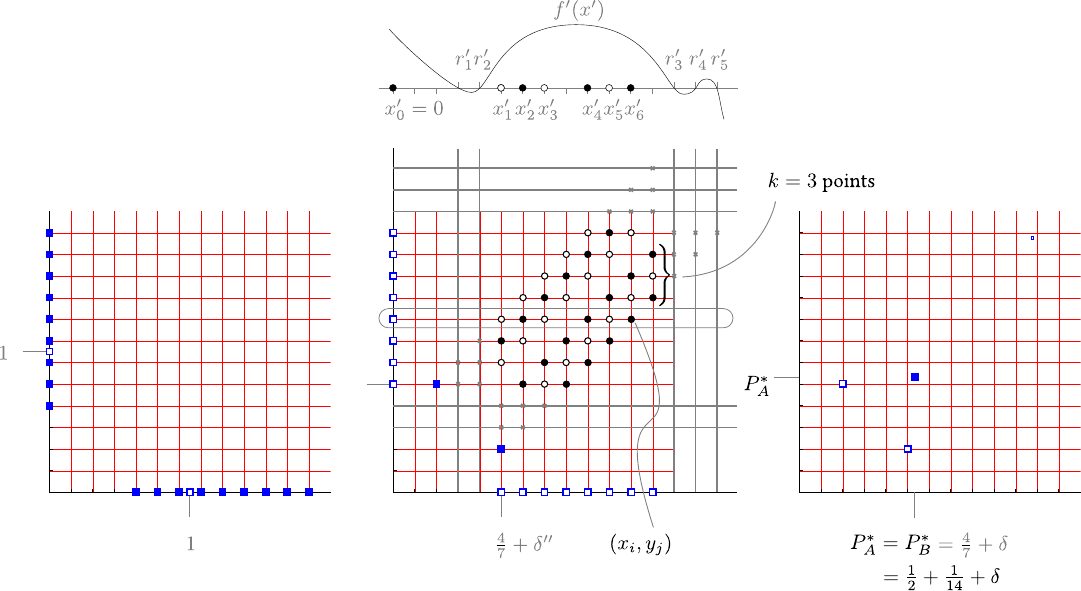}
  \par\end{centering}
  \caption{\label{fig:TDPG-1by14}The TDPG (or equivalently, the reversed protocol)
  approaching bias $\epsilon(k=3)=\frac{1}{14}$ may be seen as proceeding in
  three stages, as illustrated by the three images (left to right).
  \emph{First}, the initial points (indicated by unfilled squares) are
  split along the axes (indicated by the filled squares). \emph{Second},
  the points on the axes (unfilled squares) are transferred, via the
  ladder (indicated by the circles), into two final points (filled squares).
  \emph{Third}, the two points from the previous step (unfilled squares)
  and the catalyst state (indicated, after being raised into one point,
  by the little unfilled box) are merged into the final point (filled
  box). The \emph{second} stage is illustrated by Mochon's TIPG (or more precisely,
  the ladder) approaching bias $1/14$. Its typical move is
  highlighted. The weight of these points is given (up to a multiplicative
  constant) by the $f$\textendash assignment shown above. The roots
  of the polynomial correspond to the locations of the vertical lines
  and the location of the points in the graph is representative of the
  general construction.}
  
  \end{figure}

\subsection*{Steps of the point game}

\begin{enumerate}
\item The initial frame corresponds to the function $\frac{1}{2}\left(\left\llbracket 0,1\right\rrbracket +\left\llbracket 1,0\right\rrbracket \right)$.
\item The split move: the point $\left\llbracket 0,1\right\rrbracket $ is split
into a set of points along the $y$\textendash axis and analogously, the point
$\left\llbracket 1,0\right\rrbracket $ is split into a set of points
along the $x$\textendash axis. The number of points resulting from the splits and their respective weights match the distribution of points
along the axis as specified by the TIPG we started with.
\item The catalyst state \cite{Mochon07,Aharon2014,Arora2019}:  Deposit a small amount of weight, $\delta_{\text{catalyst}}$,
at all the points that appear in the TIPG. This can be done, for instance,
by raising (the $x$\textendash coordinates) of the points which are
along the $y$\textendash axis, i.e. if the points along the axes
are denoted as $\sum_{i}p_{\text{split},i}\left\llbracket 0,y_{i}\right\rrbracket $
then raise them to obtain $\sum_{i}(p_{\text{split},i}-\delta_{\text{split},i})\left\llbracket 0,y_{i}\right\rrbracket +\sum_{i,j}\delta_{\text{catalyst}}\left\llbracket x_{i},y_{j}\right\rrbracket $
where $\delta_{\text{catalyst}}>0$ can be chosen to be arbitrarily
small and
the second sum is over the points $(x_{i},y_{j})$ which appear in the
TIPG (excluding the points on the axes\footnote{One needs to use the analogues procedure, i.e. use $\sum_{i}p_{\text{split},i}\left\llbracket x_{i},0\right\rrbracket $
as well for the one point of the TIPG which has a $y$\textendash coordinate
smaller than that of the points along the $y$\textendash axis.}). 
\item The ladder:
\begin{enumerate}
\item The \emph{constituent functions}, i.e., the valid functions resulting from the decomposition of the valid function of the TIPG, are globally scaled such that no negative weight appears when they are applied. 
\item All the scaled down constituent horizontal functions are applied.
\item All the scaled down constituent vertical functions are applied.
\item The above two steps are repeated until all the weight has been transferred
from the axes points to the two final points of the ladder.\footnote{Once the
weight on the axes points diminishes sufficiently, it becomes impossible to apply the moves again.}
\end{enumerate}

\item The raise and merge moves: the last two points are raised and merged into the point $(1-\delta')\left\llbracket \frac{4}{7}+\delta'',\frac{4}{7}+\delta''\right\rrbracket, $
where $\delta'$ is the weight introduced by the catalyst state, and
$\delta''$ comes from the truncation of the ladder. The catalyst
state can then be absorbed (see, e.g. the proof of Theorem 5 in \cite{Aharon2014}) to obtain a single point
$\left\llbracket \frac{4}{7}+\delta,\frac{4}{7}+\delta\right\rrbracket $, 
where $\delta$ can be made arbitrarily small. 
\end{enumerate}
This final point, $\left\llbracket \frac{4}{7}+\delta,\frac{4}{7}+\delta\right\rrbracket $ with a vanishing $\delta>0$, of the point game is, in fact, the starting point of the WCF protocol. It corresponds to the initial uncorrelated state of the two parties, A and B, and the coordinates represent the cheating probabilities of each party, $P^*_{A/B}=\frac{4}{7}+\delta=\frac{1}{2}+\frac{1}{14}+\delta$. 
The steps of the point game are followed in the reverse order, and the WCF protocol ends with two points of equal weights along the axis (these are exactly the points in the initial frame of the point game) corresponding to a correlated state between A and B, $\frac{\left|00\right\rangle +\left|11\right\rangle }{\sqrt{2}}$.

\printbibliography

\end{document}